\documentclass[12pt]{article}
\usepackage{amsmath,amssymb,graphicx,calc,capt-of,ifthen,theorem}

\begin{document}

\title{A Finite Quantum Gravity Field Theory Model.}
\author{Jorge Alfaro, Pablo Gonz\'alez and Ricardo Avila\\
Pontificia Universidad Cat\'olica de Chile, Av. Vicu\~na Mackenna 4860, Santiago, Chile\\
}

\maketitle

\begin{abstract}
We discuss the quantization of Delta gravity, a two symmetric
tensors model of gravity. This model , in Cosmology, shows
accelerated expansion without a cosmological constant. We present
the $\tilde{\delta}$ transformation which defines the geometry of
the model. Then we show that all delta type models live at one
loop only. We apply this to General Relativity and we calculate
the one loop divergent part of the Effective Action showing its
null contribution in vacuum, implying a finite model. Then we
proceed to study the existence of ghosts in the model. Finally, we
study the form of the finite quantum corrections to the classical
action of the model.
\end{abstract}


\section*{INTRODUCTION.}

In the twentieth century, two major revolutions in physics changed
forever the way in which we understand nature and the world. One
of these refers to Quantum Mechanics which bares its form to
physicists like Niels Bohr, Werner Heisenberg, Erwin Schroedinger
and Paul Dirac to name a few. This framework has been used
successfully to describe the physics of the small i.e. from atoms
to quarks, their forces and interactions. The second one is
Einstein's General Theory of Relativity \cite{GR} which describes
the physics of the very large, from the motion of planets in the
solar system to the motion of galaxies in the Universe, that is,
the gravitational interaction. General Relativity, or GR for
short, is an excellent classical theory, it agrees with the
classical tests of GR \cite{SolSist} and it provides a beautiful
geometric interpretation of gravity. However, in spite of their
individual successes, there is a problem. In fact, GR is not
renormalizable. This means among other things that, we cannot
compute quantum corrections to the classical results, black hole
thermodynamics cannot be understood in statistical terms (we
cannot count states) and near the big bang, GR breaks down i.e.
quantum effects dominate the evolution of the Universe and perhaps
this could explain inflation. To solve these problems, physicist
have tried to find a unified theory that can encompass all
phenomena in all scales and, in the particular case of the
gravitational interaction, a Quantum Theory of Gravity.\\

For the first half of the past century practically no one worked
on such a thing, but in the second half, increasingly amounts of
people began to address the problem of finding a theory of quantum
gravity. The oldest approaches to quantized gravity were methods
developed by people like Gerard 'tHooft \cite{tHooft} and Bryce
DeWitt \cite{DeWitt} such as the path integral quantization and
canonical quantization respectively. Recently, the results of
these efforts have resulted in different theories which stand like
candidates to solve the problem of quantum gravity. These are
among others: superstring theory \cite{String 1} \cite{String 2},
loop quantum gravity \cite{LQG}, twistors \cite{Twistors} and
non-commutative geometry \cite{non-commutative}, but until now
none of the above have produced satisfactory results that can
single out one from the others.\\

The prime candidate for a consistent description of Quantum
Gravity is String Theory {\cite{String 1} \cite{String 2}}.
Strings appeared around 1970 to explain the confinement of quarks
inside hadrons. But later it was realized that in a closed string
theory a spin two particle that can be identified with the
graviton, naturally appears. Even in a open string theory, higher
order computations will involve closed strings as intermediate
states, so the inclusion of gravity in string theory is not only
natural but may be also unavoidable. Moreover, the finiteness of
string models make plausible the assertion that the model so
formulated provides a consistent theory of Quantum Gravity. But,
String theory has a plethora of consistent vacua, so unique
phenomenological predictions are hard to get \cite{LandScape}.\\

A second candidate to quantize gravity has developed in recent
years. It is called Loop Quantum Gravity (LQG), where a canonical
treatment of General Relativity is implemented \cite{LQG}. New
variables (loops) are introduced, in order to exploit the analogy
with Non-Abelian Yang-Mills fields. They automatically solve some
of the constraints of the theory. In a suitable Hilbert space
(spin network states) it is possible to diagonalize geometrical
objects such as volume and area, implementing its quantization.
This recent progress has permitted the computation of the Black
Hole Entropy using purely combinatoric methods \cite{bhe2}, up to
an arbitrary real number, the Barbero-Immirzi parameter. In this
formalism, the semiclassical limit is hard to get
\cite{Ashtekar}.\\

Less widely explored, but interesting possibilities, are the
search for non-trivial ultraviolet fixed points in gravity
(asymptotic safety \cite{weias}) and the notion of induced gravity
\cite{adler}. The first possibility uses exact
renormalization-group techniques \cite{litim}, and lattice and
numerical techniques such as Lorentzian triangulation analysis
\cite{loll}. Induced gravity proposed that gravitation is
a residual force produced by other interactions.\\

In a recent paper \cite{aep}, a two-dimensional field theory model
explore the emergence of geometry by the spontaneous symmetry
breaking of a larger symmetry where the metric is absent. Previous
work in this direction can be found in \cite{salam},  \cite{ogi}
and \cite{ru}.\\

Nevertheless, coming back to the original approaches, which are
still pursued today, via direct path integral quantization it is
known that, to one loop, the Einstein-Hilbert theory is, in vacuum
and without a cosmological constant, finite On-Shell \cite{tHooft}
\cite{One Loop Div}, but at two loops or more is
Non-Renormalizable \cite{Non-Renormalizable 1}
\cite{Non-Renormalizable 2} \cite{Non-Renormalizable 3}. This
means that, if we truncate the theory to one loop and evaluate in
the equations of motion, what we will have is a finite model of
gravitation. But how can we achieve this?\\

In \cite{Yang Mills} is shown a modified non abelian Yang-Mills
model, which from now on will be refer to as $\tilde{\delta}$ Yang
Mills. It is a natural and almost unique extension of non abelian
Yang Mills theory which its main characteristics are that it
preserves the classical equations of the original model at the
quantum level, introduces new symmetries, which are a natural
extension of the former symmetry, it is renormalizable, preserves
the property of asymptotic freedom and lives at one Loop in the
sense that higher loops corrections are absent. An important point
is that the quantum corrections result being the double of the
usual case.\\

Taking from the $\tilde{\delta}$ Yang Mills case and knowing the
finitness on shell for GR at one loop, the questions we raise are,
can we apply the approach of $\tilde{\delta}$ Yang Mills to GR?
Will it live at one loop? Will it give a finite model of quantum
gravity? The answer to first question is yes (see \cite{JA}) and
in the case of $\tilde{\delta}$ GR we have a natural and almost
unique extension of GR that has two tensor fields. These are the
graviton field $g_{\mu \nu}$ which transform as a two covariant
tensor under general coordinate transformations (GCT), plus
$\tilde{g}_{\mu \nu}$ which transform as a two covariant tensor
under general coordinate transformations and under an extra
symmetry. The classical aspects of the model were explored in
\cite{JA}, there it is shown that $\tilde{\delta}$ GR preserve the
classical equations of the former metric $g_{\mu\nu}$. The
equations of motion for both fields are second order, the
newtonian limit is compatible with experiments, the equivalence
principle is satisfied and, in Cosmology, accelerated expansion of
the universe is obtained without introducing a cosmological
constant.\\

In the present work, we show that all delta theories live at one
loop. We fix the gauge using the BRST method. We compute the
divergent part of the effective action and it results the double
of what was found in \cite{tHooft}. As the model lives at one loop, this
is the exact effective action. Since the equation for the original
field is preserved, this means the quantum corrections of the model
are On-Shell in the $g_{\mu \nu}$ fields so that the divergent
part of the effective action vanishes. This implies that in our
model the effective action at one loop is exact and finite
in vacuum so that it does not need to be renormalized.\\

The problem that this model has is the apparently inevitable
appearance of ghosts. Due to them, it may not be unitarity or
stable. This in turn implies difficulties with the quantization of
the model, but in \cite{phantom 1} \cite{phantom 2} \cite{phantom
3} \cite{phantom 4} \cite{phantom 5}, phantom fields are used to
explain the accelerated expansion of the universe as an
alternative to the cosmological constant and quintessence. A
feature that our model presents \cite{JA} and that our model seems
to introduce in a natural way. It would be possible that our
ghosts could be related to phantom fields in $\tilde{\delta}$ GR.
This connection may be far reaching because the phantom
idea has gained great popularity as an alternative to the cosmological constant. The present model could provide an 
arena to study the quantum properties of a phantom field, since the model has a finite quantum Effective Action.
Moreover, the advantage of being a gauge-type model, maintains
open the possibility of fixing a gauge in which the model is
unitary or impose a condition to restrict the physical Hilbert
space in such a way the model defined on this subspace is unitary.
On the other hand, as \cite{phantom 4} mentions, a choice could be
made of having either ghosts or instabilities. There the author
explains that in order to save unitary we are forced to choose
instabilities which would imply having a hamiltonian not
bounded from below.\\

Naturally, a theory of gravitation without matter is incomplete,
but it serves as a motivation for future works where the research
on these type of models can lead us to more realistic results. A
possible solution is to use $\tilde{\delta}$ Supergravity models
that contain matter fields \cite{SG} and could cure the phantom
instability.\\

On \textbf{Section \ref{Chap: delta tilde transf}} we give the
definition of $\tilde{\delta}$ transformation, we present the
general coordinate transformations and its corresponding
extensions, we define the new gauge transformations and the
generalizations of the covariant derivative. Then on
\textbf{Section \ref{Chap: Mod Theory}} we show the general form
of the invariant action for general $\tilde{\delta}$ theories,
present and demonstrate the invariance of $\tilde{\delta}$ Gravity
action and give the general form of the classical equations of
motion for general fields. In \textbf{Section \ref{Chap: Quant
correc Mod Theo}} we compute the Effective Action for a generic
$\tilde{\delta}$ model and show that all of them live at one loop.
\textbf{Section \ref{Chap: delta tilde E-H Theory}} is the most
important section of this work, here we applied what was seen in
the previous sections to the particular case of Einstein-Hilbert
theory. We show the classical equations of motion for the two
fields and give solutions for the particular cases of Schwarzchild
and Freedman-Robertson-Walker(FRW) metrics. We apply the
background field method (BFM) and give the relevant quadratic total
lagrangian. We also calculate the divergent part of the Effective
Action at one loop using an algorithm developed in \cite{One Loop
Div}. In
\textbf{Section \ref{Chap: Ghosts}}, using the gauge fixing of the
previous section, we explore the Hamiltonian formalism, redefine
the fields, the creation and annihilation operators and we see the
existence of ghosts. Finally in \textbf{Section \ref{Chap: Finite
Quantum Corrections}} we analyze the form of the finite quantum
corrections to the Effective Action and we show the modification
of the equations of motion due to the simplest type of corrections
\cite{cubic term} \cite{Dobado1} \cite{Dobado2} \cite{Non
local}.\\

In \textbf{Appendix A} we describe the gauge fixing procedure for
the modified model of GR and obtain the Faddeev-Popov for this
case using the BRST method in detail \cite{Kugo}. In
\textbf{Appendix B} we give a review of the Background Field
Method following \cite{Abbott}. Finally, in \textbf{Appendix C} we
give a brief review of the algorithm developed in \cite{One Loop
Div} for the computation of the divergent part of the Effective
Action at one loop and we indicate the values of the parameters
used in our case.\\

It is important to notice that we work with the $\tilde{\delta}$
modification to General Relativity, based on the Einstein-Hilbert
theory. From now on, we will refer to this model as
$\tilde{\delta}$ Gravity.\\

Motivated by simplicity, we will use cosmological constant
$\Lambda = 0$. We will use the Riemann Tensor given in \cite{One
Loop Div}:

\begin{eqnarray}
\label{Riemann} R^{\alpha}_{~ \beta \mu \nu} = \partial_{\mu}
\Gamma^{~~\alpha}_{\nu \beta} - \partial_{\nu}
\Gamma^{~~\alpha}_{\mu \beta} + \Gamma^{~~\alpha}_{\mu \gamma}
\Gamma^{~~\gamma}_{\nu \beta} - \Gamma^{~~\alpha}_{\nu \gamma}
\Gamma^{~~\gamma}_{\mu \beta}
\end{eqnarray}

with the Ricci Tensor $R_{\mu \nu} = R^{\alpha}_{~~ \mu \alpha
\nu}$, the Ricci scalar $R = g^{\mu \nu} R_{\mu \nu}$ and:

\begin{eqnarray}
\label{Gamma} \Gamma^{~~\alpha}_{\mu \nu} = \frac{1}{2}g^{\alpha
\beta}(\partial_{\nu} g_{\beta \mu} + \partial_{\mu} g_{\nu \beta}
- \partial_{\beta} g_{\mu \nu})
\end{eqnarray}


\section{\label{Chap: delta tilde transf}$\tilde{\delta}$
TRANSFORMATION.}

In this work we will study a modification of models that consist
in the application of a variation that we will define as
$\tilde{\delta}$. This variation will produce new elements that we
define as $\tilde{\delta}$ fields. We take through out our work
the convention that a tilde tensor is equal to the
$\tilde{\delta}$ transformation of the original tensor associated
to it when all its indexes are covariant. We raise and lower
indexes using the metric $g$.\\

In this form we will have:

\begin{eqnarray}
\label{tilde vectors} \tilde{S}_{\mu \nu \alpha ...} \equiv
\tilde{\delta}\left(S_{\mu \nu \alpha ...}\right)
\end{eqnarray}

and, for example:

\begin{eqnarray}
\tilde{\delta}\left(S^{\mu}_{~ \nu \alpha ...}\right) &=& \tilde{\delta}(g^{\mu \rho}S_{\rho \nu \alpha ...}) \nonumber \\
&=& \tilde{\delta}(g^{\mu \rho})S_{\rho \nu \alpha ...} + g^{\mu
\rho}\tilde{\delta}\left(S_{\rho \nu \alpha ...}\right)
\end{eqnarray}

It is known that $\delta(g^{\mu \nu}) = - \delta(g_{\alpha
\beta})g^{\mu \alpha}g^{\nu \beta}$, so:

\begin{eqnarray}
\label{tilde tensor} \tilde{\delta}\left(S^{\mu}_{~ \nu \alpha
...}\right) = - \tilde{g}^{\mu \rho}S_{\rho \nu \alpha ...} +
\tilde{S}^{\mu}_{~ \nu \alpha ...}
\end{eqnarray}

\subsection{\label{Sec: general coordinate transformation}General Coordinate Transformation:}

With the previous notation in mind, we can work out the general
transformations $\tilde{\delta}$ for any tensor with all its
indexes covariant (For mixed indices, please see (\ref{tilde
tensor})). We begin by considering general coordinate
transformations or diffeomorphism in its infinitesimal form:

\begin{eqnarray}
\label{diffeomorphism 0} x'^{\mu} &=& x^{\mu} - \xi_0^{\mu}(x)
\nonumber \\
\delta x'^{\mu} &=& - \xi_0^{\mu}(x)
\end{eqnarray}

Where $\delta$ is the general coordinate transformation. Now, we
define:

\begin{eqnarray}
\label{chi1}\xi_1^{\mu}(x) \equiv \tilde{\delta} \xi_0^{\mu}(x)
\end{eqnarray}

Moreover, we postulate that $\tilde{\delta}$ commutes with
$\delta$. Now we see some examples:

\subparagraph{I)} A scalar $\Phi(x)$:
\begin{eqnarray}
\label{scalar}\Phi' ( x' ) &=& \Phi ( x ) \nonumber \\
\delta \Phi ( x ) &=& \xi^{\mu}_0 \Phi_{, \mu}
 \end{eqnarray}

noting that $\tilde{\delta}$ commutes with $\delta$, we can read
the transformation rule for $\tilde{\Phi} = \tilde{\delta} \Phi$:

\begin{eqnarray}
\label{scalar_tild} \delta \tilde{\Phi} ( x ) &=& \xi^{\mu}_1
\Phi,_{\mu} + \xi^{\mu}_0 \tilde{\Phi},_{\mu}
\end{eqnarray}

\subparagraph{II)} A vector $V_{\mu} ( x )$:

\begin{eqnarray}
\label{vector} \delta V_{\mu} ( x ) = \xi_0^{\beta} V_{\mu, \beta}
+ \xi_{0, \mu}^{\alpha} V_{\alpha}
\end{eqnarray}

therefore, using (\ref{tilde tensor}), our new transformation will
be:

\begin{eqnarray}
\label{vector_tild} \delta \tilde{V}_{\mu} ( x ) = \xi_1^{\beta}
V_{\mu, \beta} + \xi_{1, \mu}^{\alpha} V_{\alpha} + \xi_0^{\beta}
\tilde{V}_{\mu, \beta} + \xi_{0, \mu}^{\alpha} \tilde{V}_{\alpha}
 \end{eqnarray}

\subparagraph{III)} Rank two Covariant Tensor $M_{\mu \nu}$:

\begin{eqnarray}
\label{tensor} \delta M_{\mu \nu} ( x ) &=& \xi^{\rho}_0 M_{\mu
\nu, \rho} + \xi_{0,\nu}^{\beta} M_{\mu \beta} + \xi_{0,
\mu}^{\beta} M_{\nu \beta}
\end{eqnarray}

and for $\tilde{M}_{\mu \nu}$,

\begin{eqnarray}
\label{tensor_tild} \delta \tilde{M}_{\mu \nu} ( x ) =
\xi^{\rho}_1 M_{\mu \nu, \rho} + \xi_{1, \nu}^{\beta} M_{\mu
\beta} + \xi_{1, \mu}^{\beta} M_{\nu \beta} + \xi^{\rho}_0
\tilde{M}_{\mu \nu, \rho} + \xi_{0, \nu}^{\beta} \tilde{M}_{\mu
\beta} + \xi_{0, \mu}^{\beta} \tilde{M}_{\nu \beta}
 \end{eqnarray}

We can define the new general coordinate transformations so that
$\delta_0$ is the transformation in $\xi_0$ and $\delta_1$ in
$\xi_1$. This new transformation is the basis of this type of
model.\\

\subsection{\label{Sec: Symm, Algebra & Gauge}Symmetry, Algebra and Gauge:}

\subsubsection{\label{SubSec: Gauge Transformation}Gauge Transformations:}

In gravitation we have a model with two fields. The first is just
the usual gravitational field $g_{\mu \nu}(x)$ and a second one
$\tilde{g}_{\mu \nu}(x)$ which corresponds to the $\tilde{\delta}$
variation of the first. We will have two gauge transformations
associated to general coordinate transformation, given by
(\ref{tensor}) and (\ref{tensor_tild}):

\begin{eqnarray}
\delta g_{\mu \nu} ( x ) &=& \xi_{0 \mu ; \nu} + \xi_{0 \nu ; \mu}
 \\
\delta \tilde{g}_{\mu \nu} ( x ) &=& \xi_{1 \mu ; \nu} + \xi_{1
\nu ; \mu} + \tilde{g}_{\mu \rho} \xi_{0, \nu}^{\rho} +
\tilde{g}_{\nu \rho} \xi^{\rho}_{0, \mu} + \tilde{g}_{\mu \nu,
\rho} \xi_0^{\rho}
\end{eqnarray}

where $\xi^{\mu}_0(x)$ and $\xi^{\mu}_1(x)$ are infinitesimal
contravariant vectors of the gauge transformations. Studying the
algebra of these transformations, we see:

\begin{eqnarray}[ \delta_{\overline{\xi}_0}, \delta_{\xi_0} ] g_{\mu \nu}( x ) = \zeta_{0 \mu ; \nu} + \zeta_{0 \nu ; \mu} = \delta_{\zeta_0} g_{\mu \nu} \end{eqnarray}

with:

\begin{eqnarray}\zeta_0^{\lambda} = \bar{\xi}^{\lambda}_{0, \rho} \xi^{\rho}_0 - \xi^{\lambda}_{0, \rho}
\bar{\xi}^{\rho}_0 \end{eqnarray}

and:

\begin{eqnarray}[ \delta_{\overline{\xi}}, \delta_{\xi} ]
\tilde{g}_{\mu \nu} ( x ) = \zeta_{1 \mu ; \nu} + \zeta_{1 \nu ;
\mu} + \tilde{g}_{\mu \rho} \zeta_{0, \nu}^{\rho} + \tilde{g}_{\nu
\rho} \zeta^{\rho}_{0, \mu} + \tilde{g}_{\mu \nu, \rho}
\zeta_0^{\rho} = \delta_{\zeta} \tilde{g}_{\mu \nu} ( x )
\end{eqnarray}

where $\zeta_0$ is as before and:

\begin{eqnarray}\zeta^{\lambda}_1 = \bar{\xi}^{\lambda}_{0, \rho} \xi^{\rho}_1 + \bar{\xi}^{\lambda}_{1, \rho} \xi^{\rho}_0 -
\xi^{\lambda}_{0,\rho} \bar{\xi}^{\rho}_1 - \xi^{\lambda}_{1,
\rho} \bar{\xi}^{\rho}_0 \end{eqnarray}

it can be seen from the above equations that both transformations
form a closed algebra.\\

\subsubsection{\label{SubSec: Covariant Differentiation}Covariant Differentiation:}

As it is usual, we define the covariant derivative as:

\begin{eqnarray}\nabla_{\nu} A_{\alpha} = D_{\nu} A_{\alpha} = A_{
\alpha ; \nu} = A_{\alpha, \nu} - \Gamma_{\alpha \nu}^{~~\lambda}
A_{\lambda} \end{eqnarray}

where $A_{\alpha}$ is a covariant vector. Now we generalize the
definition of the covariant derivative when it acts on 'tilde'
tensors e.g:

\begin{eqnarray}\nabla_{\nu} \tilde{A}_{\alpha} = \tilde{\delta} ( D_{\nu} A_{
\alpha} ) = \tilde{A}_{\alpha, \nu} - \Gamma_{\alpha
\nu}^{~~\lambda} \tilde{A}_{\lambda} -
\tilde{\delta}(\Gamma_{\alpha \nu}^{~~\lambda}) A_{\lambda}
\end{eqnarray}

where $\tilde{A}_{\alpha}=\tilde{\delta}A_{\alpha}$, and we
reserve the $D$ notation for the usual covariant derivative and
$\nabla$ for the generalized one so that:

\begin{eqnarray}\nabla_{\nu} \tilde{A}_{\alpha} = D_{\nu} \tilde{A}_{\alpha} -
\tilde{\delta}(\Gamma_{\alpha \nu}^{~~\lambda}) A_{\lambda}
\end{eqnarray}

Where:

\begin{eqnarray}
\label{conexion tilde} \tilde{\delta}(\Gamma_{\alpha
\nu}^{~~\lambda}) = \frac{1}{2}g^{\lambda
\rho}\left(D_{\nu}\tilde{g}_{\rho \alpha} +
D_{\alpha}\tilde{g}_{\nu \rho} - D_{\rho}\tilde{g}_{\alpha
\nu}\right)
\end{eqnarray}

further, the infinitesimal transformation of the modified
connection is:

\begin{eqnarray}
\label{trans conexion tilde} \delta(\tilde{\delta}\Gamma_{\mu
\nu}^{~~\varepsilon}) = \nabla_{\mu} \nabla_{\nu}
\xi_1^{\varepsilon} + R^{\varepsilon}_{~ \nu \gamma
\mu}\xi_1^{\gamma} + \tilde{\delta}(R^{\varepsilon}_{~ \nu \gamma
\mu})\xi_0^{\gamma}
\end{eqnarray}

with:

\begin{eqnarray}
\label{R tilde} \tilde{\delta}(R^{\varepsilon}_{~ \nu \gamma \mu})
= D_{\gamma} \left[\tilde{\delta}(\Gamma^{~~ \varepsilon}_{\mu
\nu})\right] - D_{\mu} \left[\tilde{\delta}(\Gamma^{~~
\varepsilon}_{\gamma \nu})\right]
\end{eqnarray}

As $D_{\nu} A_{\alpha}$ is a  two covariant tensor, $\nabla_{\nu}
\tilde{A}_{\alpha}$ is a tilde tensor of rank two and transforms
according to equation (\ref{tensor_tild}). This definition of
covariant derivative will be used in \textbf{Section \ref{Chap:
delta tilde E-H Theory}}. We notice that an analogous type of
generalization of covariant derivative was used in \cite{Yang
Mills}.\\

Now that we have established the notation and the definitions, we
can start to look for the structure of the modified models. In the
next section, we will define the new invariant action and find the
classical equations of motion.\\


\section{\label{Chap: Mod Theory}MODIFIED MODEL.}

As the general coordinate transformations were extended, we can
look for an invariant action. We start by considering a model
which is based on a given action $S_0[ \phi_I ]$ where $\phi_I$
are generic fields, then we add to it a piece which is equal to an
$\tilde{\delta}$ variation with respect to the fields and we let
$\tilde{\delta} \phi_J = \tilde{\phi}_J$ so that we have:

\begin{eqnarray}
\label{Action} S [\phi, \tilde{\phi}] = S_0 [\phi] + \kappa_2 \int
d^4x \frac{\delta S_0}{\delta \phi_I(x)}[\phi] \tilde{\phi}_I(x)
\end{eqnarray}

with $\kappa_2$ an arbitrary constant and the indexes $I$ can
represent any kind of indexes. For more details of the definition
of $\tilde{\delta}$, please see \textit{Appendix A} of \cite{JA}.
This new defined action shows the standard structure which is used
to define any modified element or function for $\tilde{\delta}$
type models, for example the gauge fixing and Faddeev Popov. Next,
we verify that this form of action is indeed the correct one for
$\tilde{\delta}$ Gravity and so is invariant to the new general
coordinate
transformation.\\

\subsection{\label{Sec: Mod. Theory Inv}The Modified Model's Invariance:}

In this paper, we will investigate the $\tilde{\delta}$ Gravity
action, obtained by the procedure sketched above:

\begin{eqnarray}
\label{action E-H mod} S[g,\tilde{g}] = \int d^{d}x
\sqrt{-g}\left(-\frac{1}{2\kappa}R + \mathcal{L}_M\right) +
\kappa_{2} \int \left(R^{\mu \nu}-\frac{1}{2}g^{\mu \nu}R-\kappa
T^{\mu\nu} \right)\sqrt{-g} \tilde{g}_{\mu \nu} d^{d}x
\end{eqnarray}

Where $\mathcal{L}_M$ is some matter lagrangian and:

\begin{eqnarray}
T^{\mu \nu} &=& - \frac{2}{\sqrt{-g}}\frac{\delta \left(\sqrt{-g}\mathcal{L}_M\right)}{\delta g_{\mu \nu}} \nonumber \\
&=& - 2 \frac{\delta \mathcal{L}_M}{\delta g_{\mu \nu}} - g^{\mu
\nu}\mathcal{L}_M
\end{eqnarray}

is the Energy Momentum Tensor. Now we must verify that
(\ref{action E-H mod}) is invariant under the following
transformations:

\begin{eqnarray}
\delta g_{\mu \nu} ( x ) &=& g_{\mu \rho} \xi_{0, \nu}^{\rho} +
g_{\nu \rho} \xi^{\rho}_{0, \mu} + g_{\mu \nu, \rho} \xi_0^{\rho}
= \xi_{0 \mu ; \nu} + \xi_{0 \nu ; \mu}
\nonumber \\
\delta \tilde{g}_{\mu \nu} ( x ) &=& \xi_{1 \mu ; \nu} + \xi_{1
\nu ; \mu} + \tilde{g}_{\mu \rho} \xi_{0, \nu}^{\rho} +
\tilde{g}_{\nu \rho} \xi^{\rho}_{0, \mu} + \tilde{g}_{\mu \nu,
\rho} \xi_0^{\rho} \nonumber
 \end{eqnarray}

We can see that (\ref{action E-H mod}) is obviously invariant
under transformations generated by $\xi_{0}^{\rho}$, since these
are general coordinate transformations and we declared
$\tilde{g}_{\mu \nu}$ to be a covariant two tensor. Under
transformations generated by $\xi_{1}^{\rho}(\delta_{1})$, $g_{\mu
\nu}$ does not change, so we have:

\begin{eqnarray}
\label{Inv Action}\delta_{1} S(g,\tilde{g}) &=& \kappa_{2} \int
\left(R^{\mu \nu}-\frac{1}{2}g^{\mu \nu}R-\kappa T^{\mu\nu}
\right)\sqrt{-g} (\delta_{1} \tilde{g}_{\mu \nu}) d^{d}x \nonumber \\
&=& \kappa_{2} \int \left(R^{\mu \nu}-\frac{1}{2}g^{\mu
\nu}R-\kappa T^{\mu\nu}
\right)\sqrt{-g}(\xi_{1\mu;\nu}+\xi_{1\nu;\mu}) d^{d}x \nonumber \\
&=& -2\kappa_{2} \int \left(R^{\mu \nu}-\frac{1}{2}g^{\mu
\nu}R-\kappa T^{\mu\nu} \right)_{;\nu}\sqrt{-g}\xi_{1\mu} d^{d}x =
0
\end{eqnarray}

We notice that we must impose conservation of the Energy-momentum
tensor $T^{\mu\nu}_{~~;\nu}=0$ so that (\ref{Inv Action}) is
fulfilled.\\

\subsection{\label{Sec: Clas Eq}Classical Equation:}

Now that we know that our action is invariant we can start to
study the model. To begin with, we will see how are the classical
equations of motion. When varied (\ref{Action}) with respect to
$\tilde{\phi}_I$, we obtain the classical equation of $\phi_I$:

\begin{eqnarray}
\label{Eq_phi} \frac{\delta S_0}{\delta \phi_I(x)}[\phi] = 0
 \end{eqnarray}

And when varied with respect to $\phi_I$, we obtain the
$\tilde{\phi}_I$'s equation:

\begin{eqnarray}
\frac{\delta S_0}{\delta \phi_I(y)}[\phi] + \kappa_2 \int d^4x
\frac{\delta^2 S_0}{\delta \phi_I(y) \delta \phi_J(x)}[\phi]
\tilde{\phi}_J(x) = 0
\end{eqnarray}

Simplifying this equation using (\ref{Eq_phi}), we obtain:

\begin{eqnarray}
\label{Eq_phi_tild} \int d^4x \frac{\delta^2 S_0}{\delta \phi_I(y)
\delta \phi_J(x)}[\phi] \tilde{\phi}_J(x) = 0
\end{eqnarray}

Where we notice that $\frac{\delta^2 S_0}{\delta \phi_J \delta
\phi_I}[\phi]$ is a differential operator acting on
$\tilde{\phi}_J$. $\tilde{\phi}_J$ belongs to the kernel of this
differential operator. It turns out that the kernel is not zero, a
fact that can be clearly seen in this paper, for the case of
gravitation, in equation (\ref{Eq grav general}) and below.\\

Having studied the classical model we can begin to look for the
quantum aspects of it. In the next section we will compute the
quantum corrections using a path integral approach.\\


\section{\label{Chap: Quant correc Mod Theo}QUANTUM MODIFIED MODEL.}

In this section, we derive the exact Effective Action for a
generic $\tilde{\delta}$ model and apply the result to the
Einstein-Hilbert action in \textbf{Section \ref{Chap: delta tilde
E-H Theory}}. We saw that the classical action for a
$\tilde{\delta}$ model is (\ref{Action}). This in turn implies
that we now have two fields to be integrated in the generating
functional of Green functions:

\begin{eqnarray}
\label{z}
  Z(j, \tilde{j}) = e^{i W(j, \tilde{j})} = \int \mathcal{D} \phi \mathcal{D}
  \tilde{\phi} e^{i \left(S_0 + \int d^N x \frac{\delta S_0}{\delta \phi_I} \tilde{\phi}_I
  + \int d^N x (j_I (x) \phi_I (x) + \tilde{j}_I (x) \tilde{\phi}_I (x))\right)}
\end{eqnarray}

We can readily appreciate that because of the linearity of the
exponent on $\tilde{\phi}_J$ what we have is the integral
representation of a Dirac's delta function so that our modified
model, once integrated, over $\tilde{\phi}_J$ gives a model with a
constrain making the original model live on-shell.

\begin{eqnarray}
\label{path_integral} Z(j, \tilde{j}) = \int \mathcal{D} \phi e^{i
\left(S_0 + \int d^N x j_I (x) \phi_I (x)\right)} \delta \left(
\frac{\delta S_0}{\delta \phi_I (x)} + \tilde{j}_I (x)\right)
\end{eqnarray}

A first glance at equation (\ref{path_integral}) could lead us to
believe that this model is purely classical. But we can see by
doing a short and simple analysis that this is not so. For this,
we follow \cite{Abers Lee}. See also \cite{ramond}.\\

Let $\varphi_I$ solves the classical equation of motion:

\begin{eqnarray}
 \frac{\delta S_0}{\delta \phi_I (x)} |_{\varphi_I} +
\tilde{j}_I (x) = 0 \label{cc}
\end{eqnarray}

We have:

\begin{eqnarray}
\delta \left( \frac{\delta S_0}{\delta \phi_I (x)} + \tilde{j}_I
(x)\right) = \det^{- 1} \left( \frac{\delta^2 S_0}{\delta \phi_I
(x) \delta \phi_J (y)} |_{\varphi_I}\right) \delta (\phi_I -
\varphi_I)
\end{eqnarray}

Therefore:

\begin{eqnarray}
  Z(j, \tilde{j}) = \int \mathcal{D} \phi e^{i (S_0 + \int d^N x j_I (x) \phi_I (x))} \delta \left(
  \frac{\delta S_0}{\delta \phi_I (x)} + \tilde{j}_I (x)\right) = e^{i (S_0 (\varphi) + \int d^N x j_I (x) \varphi_I (x))} \det^{- 1} \left(
  \frac{\delta^2 S_0}{\delta \phi_I (x) \delta \phi_J (y)}
  |_{\varphi_I}\right)
\end{eqnarray}

Notice that $\varphi$ is a functional of $\tilde{j}$. The
generating functional of connected Green functions is:

\begin{eqnarray}
  W(j, \tilde{j}) = S_0 (\varphi) + \int d^N x j_I (x) \varphi_I (x) + i
  {\rm Tr} \left(\log \left( \frac{\delta^2 S_0}{\delta \phi_I (x) \delta \phi_J (y)}
  |_{\varphi_I}\right)\right) \label{cgf}
\end{eqnarray}

Define:

\begin{eqnarray}
\Phi_I (x) &=& \frac{\delta W}{\delta j_I (x)} \nonumber \\
&=& \varphi_I (x) \nonumber \\
\tilde{\Phi}_I (x) &=& \frac{\delta W}{\delta \tilde{j}_I (x)}
\nonumber
\end{eqnarray}

The effective action is defined by:

\begin{eqnarray}
\Gamma (\Phi, \tilde{\Phi}) = W(j, \tilde{j}) - \int d^N x \left\{
j_I(x) \Phi_I (x) + \tilde{j}_I (x) \tilde{\Phi}_I (x)\right\}
\nonumber
\end{eqnarray}

We get, using equations (\ref{cc}) and (\ref{cgf}):

\begin{eqnarray}
\Gamma (\Phi, \tilde{\Phi}) = S_0 (\Phi) +  \int d^N x
\frac{\delta S_0}{\delta \Phi_I (x)} \tilde{\Phi}_I (x) + i {\rm
Tr} \left(\log \left( \frac{\delta^2 S_0}{\delta \Phi_I (x) \delta
\Phi_J (y)}\right)\right)\label{ea}
\end{eqnarray}

This is the exact effective action for $\tilde{\delta}$ theories.
In this demonstration we have assumed that all the relevant steps
for fixing the gauge have been made in (\ref{z}), so $S_0$
includes the Gauge Fixing and Faddeev Popov lagrangian, in the
manner that the new gauge fixing and Faddeev Popov will be the
original ones plus the $\tilde{\delta}$ variation of them. For
more detail on this and in the case of gravitation, see
\textbf{Appendix A} Eq. (\ref{GF grav}) and (\ref{FP grav}).\\

Comparing equation (16.42) of \cite{Abers  Lee} with equation
(\ref{ea}), we see  that the one loop contribution to the
effective action of $\delta$ theories is exact and the
$\tilde{\delta}$ modified model lives only to one loop because
higher corrections simply do not exist. Finally it is twice the
one loop contribution of the original theory from which the
$\tilde{\delta}$ model was derived. This results from having
doubled the number of degrees of freedom. We also see that this
term does not depend on the $\tilde{\phi}_I$ fields.\\

These conclusions can also be achieved through a topological
analysis with Feynman diagrams, but due to its length and
complexity we have chosen to leave out. See \cite{Yang Mills} for
the particular case of non abelian $\tilde{\delta}$ Yang Mills.\\

We see from equation (\ref{ea}) that the equations of motion for
the original field $\Phi_I (x)$ do not receive quantum
corrections:

\begin{eqnarray}
\label{EQ Quantum phi} \frac{\delta}{\delta \tilde{\Phi}_I (z)}
\Gamma (\Phi,
\tilde{\Phi}) &=& 0 \nonumber \\
\frac{\delta S_0}{\delta \Phi_I (z)} &=& 0
\end{eqnarray}

On the other side, when varying with respect to $\phi_I$, one
obtains that the equations of motion for the new field
$\tilde{\phi_I}$ do receive quantum corrections:

\begin{eqnarray}
\label{EQ Quantum phi tilde} \frac{\delta}{\delta \Phi_I (x)}
\Gamma (\Phi, \tilde{\Phi})
&=& 0 \nonumber \\
\int d^N x \frac{\delta^2 S_0}{\delta \Phi_I (z) \delta \Phi_J
(x)} \tilde{\Phi}_J (x) + i \frac{\delta}{\delta \Phi_I (z)} {\rm
Tr}\left( \log \left( \frac{\delta^2 S_0}{\delta \Phi_I (x) \delta
\Phi_J (y)}\right)\right) &=& 0
\end{eqnarray}

In conclusion, the quantum corrections behave as a source that
only affects the equations of the new field remaining the ones of
the original field unchanged. This is clearly seen when we compare
(\ref{EQ Quantum phi}) and (\ref{EQ Quantum phi tilde}) with
(\ref{Eq_phi}) and (\ref{Eq_phi_tild}).\\

In general, ${\rm Tr}\left( \log \left( \frac{\delta^2 S_0}{\delta
\Phi_I (x) \delta \Phi_J (y)}\right)\right)$ could be divergent
and need to be renormalized (See \cite{Yang Mills}). From equation
(\ref{ea}), we see that $\tilde{\delta}$ model will be
renormalizable if the original theory is renormalizable. But, due
to equation (\ref{EQ Quantum phi}), originally non-renormalizable
theories could be finite or renormalizable in the $\tilde{\delta}$
version of it. This term can be calculated in many ways, for
example by Zeta function regularization (See, for instance,
\cite{ZReg}), perturbation theory (Feynman diagrams), etc. For
gravitation, the calculation of this term is quite difficult in
any of the above methods so we will use an alternative method
developed in \cite{One Loop Div}.\\

In the present work, the $\tilde{\delta}$ Gravity model contains
two dynamical fields, $g_{\mu\nu}$ and $\tilde{g}_{\mu\nu}$, both
are important to describe the gravitational field in this
approach. Please see \cite{JA}. So, we must consider the effective
action of the model for the two fields. We saw that $g_{\mu\nu}$
satisfies the classical equations always. This is the meaning of
equation (\ref{path_integral}). But, the equation of motion for
$\tilde{g}_{\mu\nu}$ do receive quantum corrections. Moreover, One
Particle Irreducible Graphs containing $g_{\mu\nu}$ external legs
are non trivial and subjected to quantum effects.\\

In the next section, we will study $\tilde{\delta}$ Gravity. We
will see that the divergent part of the quantum corrections to the
Effective Action give a null contribution to the equations of
motion for pure gravity and without a cosmological constant, which
means that under these conditions we have a finite model of
gravity.\\


\section{\label{Chap: delta tilde E-H Theory}$\tilde{\delta}$ GRAVITY.}

Until now, we have studied $\tilde{\delta}$ models in general. We
found the invariant action given by (\ref{Action}), with the
classical equations of motion (\ref{Eq_phi}) and
(\ref{Eq_phi_tild}). Then, we demonstrated that $\tilde{\delta}$
models live only to one loop and the effective action is given by
(\ref{ea}). In this section, we apply these results to gravity. In
the first part, we will present the classical equations of motion
for both fields and show the solutions in two cases. Then we will
apply the Background Field Method (BFM) to obtain the quadratic
lagrangians and finally we calculate the divergent part of the
effective action for
$\tilde{\delta}$ Gravity.\\

\subsection{\label{Sec: New Theory} Classical Equations of Motion and Solutions:}

Now we are ready to study the modifications to gravity. In this
case, we have that $\phi_I \rightarrow g_{\mu \nu}$ and
$\tilde{\phi}_I \rightarrow \tilde{g}_{\mu \nu}$. So, using
(\ref{Action}), we obtain:

\begin{eqnarray}
\label{complete grav action}L_0[g_{\mu \nu}] &=& \sqrt{-g}\left(-\frac{1}{2\kappa}R + \mathcal{L}_M \right)\nonumber \\
L[g_{\mu \nu}] &=& \sqrt{-g}\left[-\frac{1}{2\kappa}R +
\mathcal{L}_M + \kappa_2' \left(G^{\mu \nu} - \kappa T^{\mu
\nu}\right) \tilde{g}_{\mu \nu}\right]
\end{eqnarray}

with $\kappa = \frac{8 \pi G}{c^4}$, $\kappa_2' =
\frac{\kappa_2}{2\kappa}$, $\mathcal{L}_M$ some matter lagrangian
and:

\begin{eqnarray}
\label{Tens Energ moment} T^{\mu \nu} &=& - \frac{2}{\sqrt{-g}}\frac{\delta \left(\sqrt{-g}\mathcal{L}_M\right)}{\delta g_{\mu \nu}} \nonumber \\
&=& - 2 \frac{\delta \mathcal{L}_M}{\delta g_{\mu \nu}} - g^{\mu
\nu}\mathcal{L}_M
\end{eqnarray}

is the Energy Momentum Tensor. Recall from (\ref{Inv Action}),
that we need $T^{\mu \nu}$ to be conserved. If we variate this
action, we obtain the equations of motion:

\begin{eqnarray}
\label{Eq grav general} G^{\mu \nu} &=& \kappa T^{\mu \nu} \nonumber \\
\frac{1}{2}\left(R^{\mu \nu}\tilde{g}^{\sigma}_{\sigma} - R
\tilde{g}^{\mu \nu}\right) + F^{(\mu \nu) (\alpha \beta) \rho
\lambda} D_{\rho} D_{\lambda} \tilde{g}_{\alpha \beta} &=& \kappa
\frac{\delta T_{\alpha \beta}}{\delta g_{\mu \nu}}
\tilde{g}^{\alpha \beta}
\end{eqnarray}

with:

\begin{eqnarray}
\label{F} F^{(\mu \nu) (\alpha \beta) \rho \lambda} &=& P^{((\rho
\mu) (\alpha \beta))}g^{\nu \lambda} + P^{((\rho \nu) (\alpha
\beta))}g^{\mu \lambda} - P^{((\mu \nu) (\alpha \beta))}g^{\rho
\lambda} - P^{((\rho \lambda) (\alpha \beta))}g^{\mu \nu} \nonumber \\
P^{((\alpha \beta)(\mu \nu))} &=& \frac{1}{4}\left(g^{\alpha
\mu}g^{\beta \nu} + g^{\alpha \nu}g^{\beta \mu} - g^{\alpha
\beta}g^{\mu \nu}\right)
\end{eqnarray}

Where $(\mu \nu)$ tells us that the $\mu$ and $\nu$ are in a
totally symmetric combination. An important thing to notice is
that both equations are of second order in derivatives which is
needed to preserve causality. In this paper, we will work in the
vacuum, this is $\mathcal{L}_M = 0$, so that (\ref{Eq grav
general}) simplifies to:

\begin{eqnarray}
\label{Eq grav vacuum} R^{\mu \nu} &=& 0 \nonumber \\
F^{(\mu \nu) (\alpha \beta) \rho \lambda} D_{\rho} D_{\lambda}
\tilde{g}_{\alpha \beta} &=& 0
\end{eqnarray}

Some particular solutions to equations (\ref{Eq grav general}) and (\ref{Eq grav vacuum}) are the following:\\

For the vacuum, we have for example the case of Schwarzschild:

\begin{eqnarray}
g_{\mu \nu} = \left(%
\begin{array}{cccc}
  -\left(1 - \frac{\alpha}{r}\right) &                0               &  0  &       0          \\
                  0                  & \frac{1}{1 - \frac{\alpha}{r}} &  0  &       0          \\
                  0                  &                0               & r^2 &       0          \\
                  0                  &                0               &  0  & r^2 \sin(\theta) \\
\end{array}%
\right)
\end{eqnarray}

which has a solution for $\tilde{g}_{\alpha \beta}$ of the form:

\begin{eqnarray}
\tilde{g}_{\mu \nu} = \left(%
\begin{array}{cccc}
  -\left(1 - \frac{2\alpha + \beta}{r}\right) &                                  0                              &  0  &       0          \\
                     0                        & \frac{1 + \frac{\beta}{r}}{\left(1 - \frac{\alpha}{r}\right)^2} &  0  &       0          \\
                     0                        &                                  0                              & r^2 &       0          \\
                     0                        &                                  0                              &  0  & r^2 \sin(\theta) \\
\end{array}%
\right)
\end{eqnarray}

Where it as been imposed that $g_{\mu \nu}$ and $\tilde{g}_{\mu
\nu}$ approach Minkowski space when $r\rightarrow \infty$, and
$\alpha$ and $\beta$ are determined by boundary conditions.\\

Another interesting case is for the case of
Friedman-Robertson-Walker under a density $\rho(t)$ with an
equation of state $p(t) = \omega \rho(t)$:

\begin{eqnarray}
g_{\mu \nu} = \left(%
\begin{array}{cccc}
 -1 &  0   &    0    &         0            \\
  0 & R(t) &    0    &         0            \\
  0 &  0   & R(t)r^2 &         0            \\
  0 &  0   &    0    & R(t)r^2 \sin(\theta) \\
\end{array}%
\right)
\end{eqnarray}

Where $R(t) = R_0
\left(\frac{t}{t_0}\right)^{\frac{2}{3(1+\omega)}}$. The solution
for $\tilde{g}_{\alpha \beta}$ is:

\begin{eqnarray}
\tilde{g}_{\mu \nu} = \left(%
\begin{array}{cccc}
 -\tilde{A}(r) &     0        &      0          &            0                 \\
      0        & \tilde{B}(r) &      0          &            0                 \\
      0        &     0        & \tilde{B}(r)r^2 &            0                 \\
      0        &     0        &      0          & \tilde{B}(r)r^2 \sin(\theta) \\
\end{array}%
\right)
\end{eqnarray}

with $\tilde{A}(r) = 3 \omega l_2
\left(\frac{t}{t_0}\right)^{\frac{\omega - 1}{\omega + 1}}$,
$\tilde{B}(r) = R_0^2 l_2 \left(\frac{t}{t_0}\right)^{\frac{3
\omega - 1}{3(1+\omega)}}$ and $l_2$ a free parameter. This case
was analyzed in \cite{JA}, where accelerated expansion of the
universe was obtained without a cosmological constant. This means
that $\tilde{\delta}$ Gravity does not need of dark energy, being
this one of its features that motivated us to study its
quantization.\\

\subsection{\label{Sec: BFM and Quadratic Lagrangians}BFM and
Quadratic Lagrangians:}

We proceed to calculate the quadratic lagrangians for
$\tilde{\delta}$ Gravity and Faddeev Popov. This expressions are
needed to obtain the one loop corrections of the model. For this,
we use the Background Field Method (See \textbf{Appendix B}), with
$\mathcal{L}_M = 0$. That is $g_{\mu \nu} \rightarrow g_{\mu \nu}
+ h_{\mu \nu}$ and $\tilde{g}_{\mu \nu} \rightarrow \tilde{g}_{\mu
\nu} + \tilde{h}_{\mu \nu}$. So, (\ref{complete grav action})
reduces to:

\begin{eqnarray}L_0[g_{\mu \nu} + h_{\mu \nu}] &=& \frac{\sqrt{-g}}{2\kappa}\left(-\bar{R} - \frac{1}{2}C^2\right) \nonumber \\
L[g_{\mu \nu} + h_{\mu \nu}] &=&
\frac{\sqrt{-g}}{2\kappa}\left(-\bar{R} - C^{\mu}H_{\mu} +
\kappa_2 \bar{G}^{\mu \nu} \tilde{g}_{\mu \nu}\right)
\end{eqnarray}

with $\bar{R} = R[g + h]$ and $\bar{G}^{\mu \nu} = G^{\mu \nu}[g +
h]$. We have included the original gauge fixing $C_{\mu} =
h^{\nu}_{\mu;\nu} - \frac{1}{2}h^{\nu}_{\nu;\mu}$ and the new part
$H_{\mu} = \frac{1}{2}\left(1 +
\frac{\kappa_2}{2}\tilde{g}^{\alpha}_{\alpha}\right)C_{\mu} +
\kappa_2\left( \tilde{C}_{\mu} - \frac{1}{2}\tilde{g}_{\mu \rho}
C^{\rho}\right)$ (See \textbf{Appendix A}). When we calculate the
quadratic part in the quantum gravitational fields, $h_{\mu \nu}$
and $\tilde{h}_{\mu \nu}$, we obtain:

\begin{eqnarray}
\label{L_quad} L_{quad} = \frac{1}{2}\sqrt{-g}\vec{h}^{T}_{(\alpha
\beta)}P^{((\alpha \beta)(\mu \nu))}\left(\left[K^{(\gamma
\varepsilon)}_{(\mu \nu)}\right]^{(\lambda
\eta)}\nabla_{\lambda}\nabla_{\eta} + \left[W^{(\gamma
\varepsilon)}_{(\mu \nu)}\right]\right)\vec{h}_{(\gamma
\varepsilon)}
\end{eqnarray}

and:

\begin{eqnarray}
\label{h vect}\vec{h}_{(\alpha \beta)} &=& \left(%
\begin{array}{c}
      h_{\alpha \beta}     \\
  \tilde{h}_{\alpha \beta} \\
\end{array}%
\right) \\
\label{K grav} \left[K^{(\gamma \varepsilon)}_{(\mu
\nu)}\right]^{(\lambda
\eta)} &=& \frac{1}{2\kappa}g^{\lambda \eta} \left(%
\begin{array}{cc} \left(1 + \frac{\kappa_2}{2}\tilde{g}^{\sigma}_{\sigma}\right)\delta^{\gamma \varepsilon}_{\mu \nu} + \kappa_2 P^{-1}_{((\mu \nu)(\sigma \rho))}\tilde{\delta}(P^{((\sigma \rho)(\gamma \varepsilon))}) & \kappa_2 \delta^{\gamma \varepsilon}_{\mu \nu} \\
                                                                \kappa_2 \delta^{\gamma \varepsilon}_{\mu \nu}                                                                               &                         0                     \\
\end{array}%
\right) -  \frac{\kappa_2}{2\kappa}\tilde{g}^{\lambda
\eta}\delta^{\gamma
\varepsilon}_{\mu \nu}\left(%
\begin{array}{cc}
  1 & 0 \\
  0 & 0 \\
\end{array}%
\right) \\
\label{W grav}
\left[W^{(\gamma \varepsilon)}_{(\mu \nu)}\right] &=& \frac{1}{\kappa}\left(%
\begin{array}{cc}
  \left(1 + \frac{\kappa_2}{2}\tilde{g}^{\sigma}_{\sigma}\right)X^{(\gamma \varepsilon)}_{(\mu \nu)} + \kappa_2 \tilde{\delta}(X^{(\gamma \varepsilon)}_{(\mu \nu)})  + \kappa_2 P^{-1}_{((\mu \nu)(\sigma \rho))}\tilde{\delta}(P^{((\sigma \rho)(\alpha \beta))})X^{(\gamma \varepsilon)}_{(\alpha \beta)} & \kappa_2 X^{(\gamma \varepsilon)}_{(\mu \nu)} \\
                                                                                                                  \kappa_2 X^{(\gamma \varepsilon)}_{(\mu \nu)}                                                                                                                                              &                    0                          \\
\end{array}%
\right)
\end{eqnarray}

Where:

\begin{eqnarray}
X^{(\gamma \varepsilon)}_{(\mu \nu)} = \frac{1}{2}\left(R_{\mu ~
\nu}^{~ \gamma ~ \varepsilon} + R_{\mu ~ \nu}^{~ \varepsilon ~
\gamma} +
\frac{1}{2}\left(\delta_{\mu}^{\gamma}R_{\nu}^{\varepsilon} +
\delta_{\mu}^{\varepsilon}R_{\nu}^{\gamma} +
\delta_{\nu}^{\gamma}R_{\mu}^{\varepsilon} +
\delta_{\nu}^{\varepsilon}R_{\mu}^{\gamma}\right) - \delta^{\gamma
\varepsilon}R_{\mu \nu} - \delta_{\mu \nu}R^{\gamma\varepsilon} -
\frac{1}{2}R\left(\delta_{\mu}^{\gamma}\delta_{\nu}^{\varepsilon}
+ \delta_{\mu}^{\varepsilon}\delta_{\nu}^{\gamma} - \delta_{\mu
\nu}\delta^{\gamma \varepsilon}\right)\right)
\end{eqnarray}

where $P^{((\alpha \beta)(\mu \nu))}$ is defined in (\ref{F}) and
$\delta^{\gamma \varepsilon}_{\mu \nu}$ is the symmetrized
Kronecker delta. Moreover, the covariant derivative works on
$\vec{h}_{(\gamma \varepsilon)}$ vector like:

\begin{eqnarray}
\label{Cov h}\nabla_{\lambda}\vec{h}_{(\gamma \varepsilon)} =
\partial_{\lambda}\vec{h}_{(\gamma \varepsilon)} -
\left[\Gamma^{~~\beta}_{\lambda \gamma}\right]\vec{h}_{(\beta
\varepsilon)} - \left[\Gamma^{~~\beta}_{\lambda
\varepsilon}\right]\vec{h}_{(\gamma \beta)} \end{eqnarray}

with:

\begin{eqnarray}
\label{Matrix Gamma}\left[\Gamma^{~~\beta}_{\lambda
\gamma}\right] = \left(%
\begin{array}{cc}
\Gamma^{~~\beta}_{\lambda \gamma} & 0 \\
\tilde{\delta}(\Gamma^{~~\beta}_{\lambda \gamma}) &
\Gamma^{~~\beta}_{\lambda \gamma} \\
\end{array}%
\right)
\end{eqnarray}

And using the BRST method, we obtain de Faddeev-Popov:

\begin{eqnarray}
\label{L FP} L_{FP} =
\vec{\bar{c}}_{\mu}^{T}\sqrt{-g}\left(\left[K_{FP}^{\mu
\lambda}\right]^{(\rho \nu)}\nabla_{\rho}\nabla_{\nu} +
\left[W_{FP}^{\mu \lambda}\right]\right)\vec{c}_{\lambda}
\end{eqnarray}

Where:

\begin{eqnarray}\label{c vect}\vec{c}_{\lambda} &=& \left(%
\begin{array}{c}
      c_{\lambda}     \\
  \tilde{c}_{\lambda} \\
\end{array}%
\right) \\
\label{K fant}
\left[K_{FP}^{\mu \lambda}\right]^{(\rho \nu)} &=& i g^{\nu \rho} \left(%
\begin{array}{cc}
  \frac{1}{2} \left(1 + \frac{\kappa_2}{2}\tilde{g}^{\sigma}_{\sigma}\right)g^{\mu \lambda} - \frac{\kappa_2}{2}\tilde{g}^{\mu \lambda}  & \kappa_2 g^{\mu \lambda}  \\
                        g^{\mu \lambda}                                                                                                &         0                \\
\end{array}%
\right) - i \kappa_2 \tilde{g}^{\nu \rho} g^{\mu \lambda}
\left(%
\begin{array}{cc}
  1 & 0 \\
  0 & 0 \\
\end{array}%
\right) \\
\label{W fant}
\left[W_{FP}^{\mu \lambda}\right] &=& i \left(%
\begin{array}{cc}
  \frac{1}{2}\left(1 + \frac{\kappa_2}{2}\tilde{g}^{\sigma}_{\sigma}\right)R^{\mu \lambda} - \kappa_2 \tilde{g}_{\alpha \beta}R^{\mu \alpha \lambda \beta}  - \frac{\kappa_2}{2} \tilde{g}^{\mu \alpha}R^{\lambda}_{\alpha} - g^{\alpha \beta} g^{\mu \gamma} \tilde{\delta}\left(R^{\lambda}_{~ \alpha \beta \gamma}\right) & \kappa_2 R^{\mu \lambda} \\
                                                                                                                                               R^{\mu \lambda}                                                                                                                                                          &           0             \\
\end{array}%
\right) \end{eqnarray}

with:

\begin{eqnarray}
\label{Cov ghost}\nabla_{\lambda}\vec{c}_{\mu} =
\partial_{\lambda}\vec{c}_{\mu}
- \left[\Gamma^{~~\beta}_{\lambda \mu}\right]\vec{c}_{\beta}
\end{eqnarray}

The detail of the Faddeev Popov lagrangian is presented in
\textbf{Appendix A}.\\

\subsection{\label{Sec: Divergent Part Effective Action}Divergent
Part of the Effective Action:}

In \textbf{Section \ref{Chap: Quant correc Mod Theo}} we
demonstrated that the quantum corrections to the Effective Action
do not depend on the tilde fields, in this case $\tilde{g}_{\mu
\nu}$. On the other side, Renormalization theory tells us that its
divergent corrections can only be local terms. So, by power
counting and invariance of the Background Field Effective Action
under general coordinate transformations, we know that the
divergent part to $L$ loops is \cite{tHooft}
\cite{Non-Renormalizable 3}:

\begin{eqnarray}
\label{S Loops} \Delta S^{L}_{div} \propto \int d^4x \sqrt{-g}
R^{L+1}
\end{eqnarray}

Where $R^{L+1}$ is any scalar contraction of $(L+1)$ Riemann's
tensors. As our model lives only to one loop:

\begin{eqnarray}
\label{Div Term} L^{div}_{Q} = \sqrt{-g}(a_1 R^2 + a_2 R_{\alpha
\beta}R^{\alpha \beta})
\end{eqnarray}

We do not use $R_{\alpha \beta \gamma \lambda}R^{\alpha \beta
\gamma \lambda}$ because we have the topological identity in four
dimensions:

\begin{eqnarray}
\label{Topol Iden} \sqrt{-g} \left(R_{\alpha \beta \gamma \lambda}
R^{\alpha \beta \gamma \lambda} - 4 R_{\alpha \beta} R^{\alpha
\beta} + R\right) = \textrm{Total derivative.}
\end{eqnarray}

To calculate the
divergent part of the Effective Action in our model (i.e. $a_1$ and  $a_2$ in (\ref{Div Term})), we made a FORM program \cite{FORM} to implement the algorithm developed in
\cite{One Loop Div}, obtaining in our case (\textbf{See
Appendix C}):

\begin{eqnarray}
\label{NewGravDiv}
L^{div}_{Q,grav} &=& \sqrt{-g} \frac{\hbar c}{\varepsilon}\left(\frac{7}{12}R^2 + \frac{7}{6}R_{\alpha \beta}R^{\alpha \beta}\right)\nonumber \\
L^{div}_{Q,ghost} &=& -2 \times \sqrt{-g} \frac{\hbar c}{\varepsilon}\left(\frac{17}{60}R^2 + \frac{7}{30}R_{\alpha \beta}R^{\alpha \beta}\right)\nonumber \\
L^{div}_{Q} &=& \sqrt{-g} \frac{\hbar
c}{\varepsilon}\left(\frac{1}{60}R^2 + \frac{7}{10}R_{\alpha
\beta}R^{\alpha \beta}\right) \end{eqnarray}

with $\varepsilon = 8\pi^2(N-4)$. When we compare with the usual
result in gravitation \cite{tHooft} \cite{One Loop Div} we can see
that we obtain twice the divergent term of General Relativity.
Divergences also double in Yang-Mills \cite{Yang Mills}.\\

Moreover, since Einstein's equations of motion are exactly valid
at the quantum level

\begin{eqnarray}
\label{quant eq g} \left(\frac{\delta \Gamma(g,\tilde{g})}{\delta
\tilde{g}_{\mu \nu}}\right) = R^{\mu \nu} = 0
\end{eqnarray}

Where $\Gamma(g,\tilde{g})$ is the Effective Action in the
Background Field Method. It follows that the contribution of
(\ref{NewGravDiv}) to the equation of motion vanishes:

\begin{eqnarray}
\label{grav finite} \hbar
c\left[\frac{\sqrt{-g}}{\varepsilon}\left(\frac{1}{2}g^{\mu
\nu}\left(\frac{1}{60}R^2 + \frac{7}{10}R_{\alpha \beta}R^{\alpha
\beta}\right) + \frac{1}{30}R \frac{\delta R}{\delta g_{\mu \nu}}
+ \frac{7}{10}R_{\alpha \beta}\frac{\delta R^{\alpha
\beta}}{\delta g_{\mu \nu}} + \frac{7}{10}R^{\alpha
\beta}\frac{\delta R_{\alpha \beta}}{\delta g_{\mu
\nu}}\right)\right]_{R_{\alpha \beta} = 0} = 0
\end{eqnarray}

Therefore, $\tilde{\delta}$ Gravity is a finite model of
gravitation if we do not have matter and cosmological constant.
The finiteness of our model implies that Newton Constant does not
run at all, nor with time or energy scale which would be supported
by the very stringent experimental bounds set on its change
\cite{NCGI} \cite{NCGII}. We must notice that this model is finite
only in four dimensions because we need (\ref{Topol Iden}).
Moreover, in more dimensions there could appear more terms in
(\ref{Div Term}) that contains $R^{\mu_1 \mu_2 ... \mu_N}$ with
$N$ the dimension of space, that give a non zero contribution to
the equations of motion.\\

In spite of these apparent successes there seems to be a problem
with this model and this is the possible existence of ghosts. This
issue will be dealt with in next section.\\


\section{\label{Chap: Ghosts} GHOSTS.}

In this section we discuss the possibility that our model has
ghosts and the lost of unitarity due to them. In order to proceed
with this endeavor we first write the quadratic Lagrangian
(\ref{L_quad}) for a non interacting model (this is, with the
backgrounds both equal to the Minkowsky metric tensor) and
calculate from it the canonical conjugated momenta to the quantum
fields. It is important to notice that for the Lagrangian
(\ref{L_quad}) a gauge has been chosen. Then it is possible to
show that under these conditions and in this gauge, the quantum
fields obey the wave equation and an expansion in plane waves is
possible where the Fourier coefficients are promoted to creation
and annihilation operators much in the same way as can be done for
the electromagnetic potential. We use the canonical commutation
relations for fields and momenta to work out the corresponding
canonical commutation relations for the creation and annihilation
operators. We also show first the Hamiltonian in terms of fields
and momenta and then in terms of annihilation and creation
operators.\\

To study the existence of ghosts in the model we will study small
perturbations to flat space. This is done by taking expression
(\ref{L_quad}) and putting the backgrounds equal to the minkowski
metric $g_{\mu \nu} = \eta_{\mu \nu}$ and $\tilde{g}_{\mu \nu} =
\eta_{\mu \nu}$, thus obtaining:

\begin{eqnarray}
\label{Action simply gauge}
S[h,\tilde{h}] = -\frac{1}{2\kappa}\int
d^4x P^{((\alpha \beta)(\mu
\nu))}\left(\frac{(1-\kappa_2)}{2}\partial_{\rho}h_{\alpha
\beta}\partial^{\rho}h_{\mu \nu} + \kappa_2
\partial_{\rho}\tilde{h}_{\alpha \beta}\partial^{\rho}h_{\mu
\nu}\right)
\end{eqnarray}

where now:

\begin{eqnarray}
P^{((\alpha \beta)(\mu \nu))} &=& \frac{1}{4}\left(\eta^{\alpha
\mu}\eta^{\beta \nu} + \eta^{\alpha \nu}\eta^{\beta \mu} -
\eta^{\alpha \beta}\eta^{\mu \nu}\right)
\end{eqnarray}

and the equations of motion for the fields are:

\begin{eqnarray}
\label{wave eqn h}
\partial^2 h_{\mu \nu} &=& 0 \nonumber \\
\partial^2 \tilde{h}_{\mu \nu} &=& 0
\end{eqnarray}

With $\partial^2 = \eta^{\rho
\lambda}\partial_{\rho}\partial_{\lambda}$. This corresponds to
the wave equation with energy $E_{\mathbf{p}} = |\mathbf{p}|$.
Here we notice that in order to obtain these equations, we have
made use of a particular gauge fixing term (\ref{Gauge Fix.}) in
the Lagrangian (\ref{L_quad}).\\

It is well known that for diffeomorfism invariant Lagrangian, the
canonical Hamiltonian is zero. This is so in delta-gravity as well
as in General Relativity: the total Hamiltonian is a linear
combination of the first class constraints (See \cite{DeWitt}).
After gauge fixing, the Hamiltonian is:

\begin{eqnarray}
H = \int d^3x \left(\frac{2\kappa}{\kappa_2}P^{-1}_{((\alpha
\beta)(\mu \nu))}\left(\tilde{\Pi}^{\alpha \beta}\Pi^{\mu \nu} -
\frac{(1-\kappa_2)}{2\kappa_2}\tilde{\Pi}^{\alpha
\beta}\tilde{\Pi}^{\mu \nu}\right) +
\frac{\kappa_2}{2\kappa}P^{((\alpha \beta)(\mu
\nu))}\left(\partial_i \tilde{h}_{\alpha \beta} \partial_i h_{\mu
\nu} + \frac{(1-\kappa_2)}{2\kappa_2} \partial_i h_{\alpha \beta}
\partial_i h_{\mu \nu}\right)\right)
\end{eqnarray}

with:

\begin{eqnarray}
P^{-1}_{((\alpha \beta)(\mu \nu))} = \eta_{\alpha \mu}\eta_{\beta
\nu} + \eta_{\alpha \nu}\eta_{\beta \mu} - \eta_{\alpha
\beta}\eta_{\mu \nu} = 4P_{((\alpha \beta)(\mu \nu))}
\end{eqnarray}

and where the conjugate momenta are:

\begin{eqnarray}
\Pi^{\mu \nu} &=& \frac{\delta \mathcal{L}}{\delta \dot{h}_{\mu \nu}} \nonumber \\
&=& \frac{1}{2\kappa}P^{((\alpha \beta)(\mu
\nu))}\left((1-\kappa_2)\dot{h}_{\alpha \beta} +
\kappa_2\dot{\tilde{h}}_{\alpha \beta}\right) \\
\tilde{\Pi}^{\mu \nu} &=& \frac{\delta \mathcal{L}}{\delta \dot{\tilde{h}}_{\mu \nu}} \nonumber \\
&=& \frac{\kappa_2}{2\kappa}P^{((\alpha \beta)(\mu \nu))}
\dot{h}_{\alpha \beta}
\end{eqnarray}

We can write our fields $h$ y $\tilde{h}$ the following way:

\begin{eqnarray}
\label{h htild} h_{\mu \nu}(\mathbf{x}, t) &=& \int
\frac{d^3p}{\sqrt{(2\pi)^32E_{\mathbf{p}}}}\left[\chi^{(A
B)}_{(\mu \nu)}(\mathbf{p})a_{(A B)}(\mathbf{p})e^{ip \cdot x} +
\chi^{(A B)}_{(\mu
\nu)}(\mathbf{p})a^+_{(A B)}(\mathbf{p})e^{-ip \cdot x}\right]|_{p_0 = E_{\mathbf{p}}} \nonumber \\
\tilde{h}_{\mu \nu}(\mathbf{x}, t) &=& \int
\frac{d^3p}{\sqrt{(2\pi)^32E_{\mathbf{p}}}}\left[\chi^{(A
B)}_{(\mu \nu)}(\mathbf{p})\tilde{a}_{(A B)}(\mathbf{p})e^{ip
\cdot x} + \chi^{(A B)}_{(\mu \nu)}(\mathbf{p})\tilde{a}^+_{(A
B)}(\mathbf{p})e^{-ip \cdot x}\right]|_{p_0 = E_{\mathbf{p}}}
\end{eqnarray}

where $\chi^{(A B)}_{(\mu \nu)}(\mathbf{p})$ is a polarization
tensor and $a_{(A B)}(\mathbf{p})$ and $\tilde{a}_{(A
B)}(\mathbf{p})$ are promoted to annihilation operators when we
quantize it. $a^{+}_{(A B)}(\mathbf{p})$ and $\tilde{a}^{+}_{(A
B)}(\mathbf{p})$ correspond to the creation operators. $A$ and $B$
are indices of polarization that work like Lorentz indices, this
is, they go from $0$ to $3$ and are moved up and down with
$\eta^{AB}$. As this indices are presented symmetrically we will
have ten polarization tensors, enough to make a complete basis.
For quantization of the model, we must impose the canonical
commutation relations, the only non vanishing commutators are:

\begin{eqnarray}
[h_{\mu \nu}(t,\mathbf{x}), \Pi^{\alpha \beta}(t,\mathbf{y})] =
[\tilde{h}_{\mu \nu}(t,\mathbf{x}), \tilde{\Pi}^{\alpha
\beta}(t,\mathbf{y})] = i \delta^{\alpha \beta}_{\mu \nu}
\delta^3(\mathbf{x} - \mathbf{y})
\end{eqnarray}

when expressed using (\ref{h htild}) the non-vanishing commutators
are:

\begin{eqnarray}
\label{commu1}
[a^{A B}(\mathbf{p}), \tilde{a}^+_{C
D}(\mathbf{p}')] = [\tilde{a}^{A B}(\mathbf{p}), a^+_{C
D}(\mathbf{p}')] = \frac{4 \kappa}{\kappa_2} \delta^{A B}_{C D}
\delta^3(\mathbf{p} - \mathbf{p}')
\end{eqnarray}

\begin{eqnarray}
\label{commu2}
[\tilde{a}^{A B}(\mathbf{p}), \tilde{a}^+_{C
D}(\mathbf{p}')] = -\frac{4 \kappa (1-\kappa_2)}{\kappa_2^2}
\delta^{A B}_{C D} \delta^3(\mathbf{p} - \mathbf{p}')
\end{eqnarray}

there is a slight subtlety in calculating the above commutators.
Basically the expression that appears at one stage of the calculus
is:

\begin{eqnarray}
\sum_{ABCD}\chi^{(AB)}_{(\mu\nu)}P^{(\alpha\beta)}_{(\gamma\epsilon)}\chi^{(\gamma\epsilon)}_{CD}
=
\sum_{ABCD}\chi^{(AB)}_{(\mu\nu)}\frac{1}{2}\delta^{(\alpha\beta)}_{(\gamma\epsilon)}\chi^{(\gamma\epsilon)}_{CD}
-\frac{1}{4}\eta^{\alpha\beta}\chi^{(AB)}_{(\mu\nu)}Tr(\chi)
\end{eqnarray}

and since we have the completeness relation:

\begin{eqnarray}
\sum_{ABCD}\chi^{(AB)}_{(\mu\nu)}\chi^{(\alpha\beta)}_{(CD)}\delta^{(CD)}_{(AB)}=\delta^{(\alpha\beta)}_{(\mu\nu)}
\end{eqnarray}

we must impose $Tr(\chi)=0$ which in turn means that
$Tr(h)=Tr(\tilde{h})=0$. This can always be done because the Gauge
fixing being used does not fix entirely the gauge freedom and this
further condition can be imposed (See \cite{GMTW}).\\

The Hamiltonian expressed in terms of creation and annihilation
operators is:

\begin{eqnarray}
\label{H(a)} H = \int \frac{d^3p}{4 \kappa}E_{\mathbf{p}}
\left((1-\kappa_2)a^+_{A B}a^{A B} + \kappa_2 a^+_{A
B}\tilde{a}^{A B} + \kappa_2 \tilde{a}^+_{A B}a^{A B}\right)
\end{eqnarray}

where we have subtracted an infinite constant. Looking at this
Hamiltonian we notice that it has cross products of operators,
which obscures its physical interpretation. Something analogous
happens when we observe the commutators (\ref{commu1}) and
(\ref{commu2}) and so it is difficult to define their action over
states. Because of this is that we redefine our annihilation (and
therefore also the creation) operators, for this, we return to our
action (\ref{Action simply gauge}) and define:

\begin{eqnarray}
\label{redef h}
h_{\mu\nu} &=& A\bar{h}^{1}_{\mu\nu}+B\bar{h}^{2}_{\mu\nu} \nonumber \\
\tilde{h}_{\mu\nu} &=& C\bar{h}^{1}_{\mu\nu}+D\bar{h}^{2}_{\mu\nu}
\end{eqnarray}

where A, B, C and D are real constants so that the new fields,
$\bar{h}^1$ and $\bar{h}^2$, are real fields. When replacing this
in (\ref{Action simply gauge}) we obtain:

\begin{eqnarray}
\label{S h1h2} S[\bar{h}^{1},\bar{h}^{2}] &=&
\frac{1}{2\kappa}\int d^4x P^{((\alpha \beta)(\mu
\nu))}\left(\frac{A}{2}(A-\kappa_2A+2\kappa_2C)\bar{h}^{1}_{\alpha
\beta}\partial^{2}\bar{h}^{1}_{\mu \nu} +
\frac{B}{2}(B-\kappa_2B+2\kappa_2D) \bar{h}^{2}_{\alpha
\beta}\partial^{2}\bar{h}^{2}_{\mu
\nu}\right) \nonumber \\
&& +
P^{((\alpha\beta)(\mu\nu))}(AB-\kappa_2AB+\kappa_2AD+\kappa_2BC)\bar{h}^{1}_{\alpha
\beta}\partial^{2}\bar{h}^{2}_{\mu\nu}
\end{eqnarray}

with the objective of decoupling the new fields, we make null the
last term in (\ref{S h1h2}). It can be demonstrated that imposing
the above criteria, it is inevitable that one (and only one) of
two fields will be a ghost. We make the choice of $\bar{h}^2$ as
the corresponding ghost. Taking the above considerations plus the
condition that (\ref{S h1h2}) to have the usual form of an action
with real fields, we impose that the coefficients of the first and
second terms in it to be $\frac{1}{2}$ and $-\frac{1}{2}$
respectively. This mean:

\begin{eqnarray}
A&=&B\nonumber\\
C&=&\frac{1-(1-\kappa_2)B^2}{2\kappa_2B}\nonumber\\
D&=&-\frac{1+(1-\kappa_2)B^2}{2\kappa_2B}
\end{eqnarray}

where $B$ is left as an arbitrary real constant. Here we make the
point that if we had chosen $\bar{h}^1$ as the ghost then, the
real constants change such that $C\leftrightarrow D$.\\

Then, the action we are left finally is:

\begin{eqnarray}
\label{S h1h2II} S[\bar{h}^{1},\bar{h}^{2}] =
\frac{1}{2\kappa}\int d^4x P^{((\alpha \beta)(\mu
\nu))}\left(\frac{1}{2}\bar{h}^{1}_{\alpha
\beta}\partial^{2}\bar{h}^{1}_{\mu \nu}
-\frac{1}{2}\bar{h}^{2}_{\alpha \beta}\partial^{2}\bar{h}^{2}_{\mu
\nu}\right)
\end{eqnarray}

Following this same line of reasoning we can find the destruction
operators for $\bar{h}^1$ and $\bar{h}^2$:

\begin{eqnarray}
b^1_{AB}(\vec{p})&=&\frac{1+B^2(1-\kappa_2)}{2B}a_{AB}(\vec{p})+\kappa_2B\tilde{a}_{AB}(\vec{p})\\
b^2_{AB}(\vec{p})&=&\frac{1-B^2(1-\kappa_2)}{2B}a_{AB}(\vec{p})-\kappa_2B\tilde{a}_{AB}(\vec{p})
\end{eqnarray}

where we have used (\ref{redef h}). It can be verified that the
only non vanishing commutators are now:

\begin{eqnarray}
[b^{1(AB)}(\vec{p}),b^{1+}_{CD}(\vec{p'})] &=& 4\kappa
\delta^{AB}_{CD} \delta^3(\vec{p}-\vec{p'})
\end{eqnarray}

\begin{eqnarray}
[b^{2(AB)}(\vec{p}),b^{2+}_{CD}(\vec{p'})] &=& -4\kappa
\delta^{AB}_{CD} \delta^3(\vec{p}-\vec{p'})
\end{eqnarray}

These commutators indicate that $b^1$ and $b^2$ have a vanishing
inner product and that $b^2$ is the annihilation operator for the
ghost. On the other hand, the Hamiltonian expressed in terms of
these operators is:

\begin{eqnarray}
H=\int \frac{d^3p}{4\kappa}
E_\mathbf{p}(b^{1+}_{AB}b^{1AB}-b^{2+}_{AB}b^{2AB})
\end{eqnarray}

Due to the existence of the ghost is possible that this model will
not be unitarity. To analyze this in greater depth it is necessary
to do a more profound study of the $S$-Matrix, but to do this for
gravitation is a colossal task that would takes us beyond the
original scope of this paper. On the other side, the existence of
ghost or phantom fields have been proposed by some authors to
explain the accelerated expansion of the universe \cite{phantom 1}
\cite{phantom 2} \cite{phantom 3} \cite{phantom 4} \cite{phantom
5} a feature that our model presents \cite{JA}. The problem with
these models are that when they are quantized either there is a
lost of unitarity or there are negative energy which mean lost of
stability. Looking at (\ref{S h1h2II}) we find that the
propagators of $\bar{h}^1$ and $\bar{h}^2$ are respectively:

\begin{eqnarray}
-2\kappa P^{-1}_{((\alpha \beta)(\mu \nu))}\frac{i}{p^2-i\varepsilon} \\
2\kappa P^{-1}_{((\alpha \beta)(\mu \nu))}\frac{i}{p^2\pm
i\varepsilon}
\end{eqnarray}

where $\pm$ in the phantom propagator, $\bar{h}^2$, will decide
whether unitarity and negative energy solutions or nonunitary and
positive energy solutions will be present in the model
\cite{phantom 4}.\\

The advantage that our model has against other models that use
scalar fields for the phantoms is that being a gauge model, there
remains open the possibility of fixing a gauge in which the model
is unitary keeping the model good attributes, as in the BRST
canonical quantization \cite{Kugo 2}. On the other hand as
possible solution to case of instability, we may consider
$\tilde{\delta}$ Supergravity which may solve the unboundedness
from below of the Hamiltonian. The last argument comes from the
fact that in supersymmetry one defines the hamiltonian as the
squared of an hermitian charge, making it positive definite
\cite{SusyHam1} \cite{SusyHam2}.\\

Having explained the problem that our model has, now we would like
to discuss the new physics that our model may predict. For this,
we will analyze the type of some finite quantum corrections and
how the most simplest ones affect the equations of motion of
the model.\\


\section{\label{Chap: Finite Quantum Corrections}Finite Quantum
Corrections.}

The finite quantum corrections to our modified model of gravity
can be separated into two groups. The first are the non-local
terms, which are characterized by the presence of a logarithm, in
the form \cite{Non local}:

\begin{eqnarray}
\label{Non Local term} \sqrt{-g} R_{\mu \nu} \ln \left(
\frac{\nabla^2}{\mu^2} \right) R^{\mu \nu}
\nonumber\\
\sqrt{-g} R \ln \left( \frac{\nabla^2}{\mu^2} \right) R
\end{eqnarray}

where $\nabla^2 = g^{\alpha \beta} \nabla_{\alpha}
\nabla_{\beta}$, $\nabla_{\beta}$ being the covariant derivative.
There are no terms like the above ones but quadratic in the
Riemann tensor because these terms always occurs like:

\begin{eqnarray}
\frac{1}{\epsilon} + \ln \left( \frac{\nabla^2}{\mu^2} \right)
\end{eqnarray}

and it is known that the terms that appear with the pole are
purely Ricci tensors and Ricci scalars \cite{tHooft} \cite{One
Loop Div} (see eq. (\ref{NewGravDiv}) too), which in turn is due
to (\ref{Topol Iden}). Now, when looking at the quantum
corrections and Eq. (\ref{quant eq g}), we need to care about the
variations of (\ref{Non Local term}) with respect to $g_{\mu
\nu}$. Taking this into consideration, for the non local terms we
have:

\begin{eqnarray}
\delta \left(\sqrt{-g}\right) R_{\mu \nu} \ln \left(
\frac{\nabla^2}{\mu^2} \right) R^{\mu \nu}
&=& 0 \nonumber \\
\sqrt{-g} R_{\mu \nu} \delta \left( \ln \left(
\frac{\nabla^2}{\mu^2}
\right) R^{\mu \nu} \right) &=& 0 \nonumber \\
\sqrt{-g} \delta ( R_{\mu \nu} ) \ln \left( \frac{\nabla^2}{\mu^2}
\right)R^{\mu \nu} &=& 0 \nonumber \\
\delta \left(\sqrt{-g}\right) R \ln \left( \frac{\nabla^2}{\mu^2} \right) R &=& 0 \nonumber \\
\sqrt{-g} R \delta \left(  \ln \left( \frac{\nabla^2}{\mu^2}
\right) R \right) &=& 0 \nonumber \\
\sqrt{-g} \delta ( R ) \ln \left( \frac{\nabla^2}{\mu^2} \right) R
&=& 0
\end{eqnarray}

because our model lives on shell, i.e. $R_{\mu \nu} \equiv 0$ and
$R \equiv 0$. So we see that the only relevant quantum corrections
will come from the second group, that is, from the local terms
which corresponds to a series expansion in powers of the curvature
tensor. The linear term is basically $R$, which corresponds to the
original action, and the quadratic terms when taking into account
their contribution is null due to (\ref{Topol Iden}). The next
terms to consider are cubic in the Riemann tensor. In principle
any power of the curvature tensor will appear, but we want to
discuss now only the cubic ones because they are the simpler to be
dealt with \cite{cubic term}. The most general form of these
corrections is:

\begin{eqnarray}
\label{Lfin_Q} L^{fin}_Q = \sqrt{-g} \left( c_1 ~ R_{\mu \nu
\lambda \sigma} R^{\alpha \beta \lambda \sigma} R^{\mu \nu}_{~ ~
\alpha \beta} ~ + ~ c_2 ~ R^{\mu \nu}_{~ ~ \lambda \sigma} R_{\mu
\alpha}^{~ ~ \lambda \beta} R^{\alpha \sigma}_{~ ~ \nu \beta} ~ +
~ c_3 ~ R_{\mu \nu} R^{\mu \alpha \beta \gamma} R^{\nu}_{~ \alpha
\beta \gamma} ~ + ~ c_4 ~ R R_{\mu \nu \lambda \kappa} R^{\mu \nu
\lambda \kappa} \right) \end{eqnarray}

These type of corrections will affect the equations of motion for
$\tilde{g}_{\mu \nu}$. So, using (\ref{EQ Quantum phi tilde}) we
obtain:

\begin{eqnarray}
\label{gtilde eq grav}F^{(\mu \nu) (\alpha \beta) \rho
\lambda}D_{\rho}D_{\lambda}\tilde{g}_{\alpha \beta} = -
\frac{1}{\kappa_2} \left(M^{(\mu \nu)} + c_1 N^{(\mu \nu)} + c_2
B^{(\mu \nu)} + 3\left\{D_{\rho}\textrm{ ,
}D{\sigma}\right\}E^{[\sigma \mu][\nu \rho]}\right)
\end{eqnarray}

with:

\begin{eqnarray}
\label{M} M^{(\mu \nu)} &=&
\frac{1}{2}\left(D_{\alpha}D^{\nu}A^{(\alpha \mu)} +
D_{\alpha}D^{\mu}A^{(\alpha \nu)} - D_{\alpha}D^{\alpha}A^{(\mu
\nu)} - g^{\mu
\nu}D_{\alpha}D_{\beta}A^{(\alpha \beta)}\right) \\
\label{A} A^{(\mu \nu)} &=& c_3 R^{\mu \alpha \beta \gamma}
R^{\nu}_{~ \alpha \beta \gamma} + c_4 g^{\mu \nu} R^{\alpha \beta
\gamma
\epsilon} R_{\alpha \beta \gamma \epsilon} \\
\label{N} N^{(\mu \nu)} &=& \frac{1}{2}g^{\mu \nu} R_{\rho
\epsilon \lambda \sigma} R^{\lambda \sigma \alpha \beta} R_{\alpha
\beta}^{~ ~ ~ \rho \epsilon} + 3 R_{\rho \epsilon \lambda \sigma}
R_{\alpha}^{~
\nu \epsilon \rho} R^{\alpha \mu \lambda \sigma} \\
\label{B} B^{(\mu \nu)} &=& \frac{1}{2}g^{\mu \nu} R_{\rho
\epsilon \lambda \sigma} R^{\rho \alpha \lambda \beta} R_{\alpha ~
~ \beta}^{~ \sigma \epsilon} + 3 R_{\rho \epsilon \lambda \sigma}
R^{\nu
\sigma \rho}_{~ ~ ~ \beta} R^{\mu \epsilon \beta \lambda} \\
\label{E} E^{[\sigma \mu][\nu \rho]} &=& c_1 R^{\sigma \mu}_{~ ~
\alpha \beta} R^{\alpha \beta \nu \rho} +
\frac{1}{2}c_2\left(R^{\nu ~ \sigma}_{~ \alpha ~ \beta} R^{\rho
\beta \alpha \mu} - R^{\rho ~ \sigma}_{~ \alpha ~ \beta} R^{\nu
\beta \alpha \mu}\right)
\end{eqnarray}

where $[\mu \nu]$ that $\mu$ and $\nu$ are in a antisymmetric
combination, and $F^{(\mu \nu) (\alpha \beta) \rho \lambda}$ was
defined in (\ref{F}). Obviously, if we do not have Quantum
correction, i.e: $c_1 = c_2 = c_3 = c_4 = 0$, (\ref{gtilde eq
grav}) is transformed in (\ref{Eq grav vacuum}). It is possible to
demonstrate that one solution to (\ref{Eq grav vacuum}) is
$\tilde{g}_{\mu \nu} = g_{\mu \nu}$, a fact that is necessary so
that the predictions of the original theory of Einstein-Hilbert
still are fulfilled in vacuum. This means, the solution of
(\ref{gtilde eq grav}) must come to be small perturbations to
$g_{\mu \nu}$.\\

$\tilde{\delta}$ Gravity will provide finite answers for the
constants $c_i$. Due to the general structure of the finite
quantum corrections, they will be relevant only at very short
distances and strong curvatures. So the natural scenario to test
the predictions of the model is the Inflationary Epoch of The
Universe. The computation of the $c_i$ and the phenomenological
implications of Quantum $\tilde{\delta}$ Gravity will be discussed
elsewhere.\\


\section*{CONCLUSIONS.}

We have shown following \cite{Yang Mills} that the
$\tilde{\delta}$ transformation, applied to any theory, produce
physical models that live only at one loop. This is achieved
introducing new fields that generate a new constrain through a
functional Dirac's delta inside the path integral
(\ref{path_integral}). We have seen that the original symmetries
are generalized when we apply the $\tilde{\delta}$ transformation.
Moreover, the modified model is invariant under the generalized
symmetries.\\

Now, going to $\tilde{\delta}$ Gravity we calculated the divergent
part of the action to one loop and we obtained twice the well
known result of \cite{tHooft}. We see that this factor of two
appears also in \cite{Yang Mills}. The divergent part at one loop
is zero in the absence of matter and on shell, so $\tilde{\delta}$
Gravity is a finite quantum model,  in four dimensional
space-time. This in turn implies that Newton Gravitational
Constant does not run with scale, which agrees with the very
stringent experimental bounds that restrict its variation
\cite{NCGI} \cite{NCGII}.\\

We have shown that perturbing around the Minkowsky vacuum and
using a particular Lorentz invariant gauge, we can redefine the
gravitational fields in such a way that the free part of the
action is decoupled. In this redefinition it is seen that one of
the new fields is a ghost. In spite of that this may bring unitary
or unstable problems (negative energies), these ghosts (phantoms)
can explain at a classical level the accelerated expansion of the
universe \cite{JA}. Scalar phantoms has been introduced in order
to explain Dark Energy in  \cite{phantom 1}  and discussed in many
papers. See for instance, \cite{phantom 2} \cite{phantom 3}
\cite{phantom 4} \cite{phantom 5}. 
This connection may be far reaching because the phantom
idea has gained great popularity as an alternative to the cosmological constant.The present model could provide an 
arena to study the quantum properties of a phantom field, since the model has a finite quantum Effective Action.
In this respect, the advantage
of the present model is that, being a gauge model, could give us
the possibility to solve the problem of lack of unitarity using
standard techniques of gauge theories as the BRST method. This is
something that need to be studied further but go beyond the
original scope of this paper.\\ 

We want to point out that
Supergravity with matter is finite at one Loop level \cite{SG}.
According to the general argument developed in this paper,
$\tilde{\delta}$ Supergravity will be a one Loop model which have
a strong possibility to be a finite quantum model of gravity plus
matter and also it may solve the instability of negative energies
since in supersymmetry one has an hermitian charge whose square is
equal to the Hamiltonian operator meaning that the Hamiltonian is
positive definite \cite{SusyHam1} \cite{SusyHam2}.\\

Finally we have shown that the contribution of quadratic local and
non-local logarithmic terms are zero due to the on-shell condition
of the modified model. We have also shown how the cubic
corrections in the Riemann tensor affect the equation of motion
(\ref{gtilde eq grav}). Given the general form of the quantum
corrections in quantum $\tilde{\delta}$ Gravity, they may be
important during the Inflationary Epoch of the Universe.\\



\section*{Acknowledgments}

R. A.  wants to thank partial funding given by the Department of
Physics at  PUC and to the VRAID scholarship of PUC.PG
acknowledges support from Beca Doctoral Conicyt \# 21080490.The
work of JA is partially supported by VRAID/DID/46/2010 and
Fondecyt 1110378. Special thanks are given to J. Gamboa for
carefully reading of the manuscript.

\newpage


\section*{APPENDIX A: BRST Formalism.}

First we give the BRST transformations $\bar{\delta}$ of our
model:

\begin{eqnarray}
\xi_0^{\mu} ( x ) &=& \lambda c_0^{\mu} ( x ) \nonumber \\
\xi_1^{\mu} ( x ) &=& \lambda c_1^{\mu} ( x )
\end{eqnarray}

where $\lambda$ is a Grassmann constant and $c_0^{\mu}$,
$c_1^{\mu}$ are the two ghosts of our model. Starting from the
gauge transformations for our quantum fields $h_{\mu \nu}$ and
$\tilde{h}_{\mu \nu}$ \cite{Weinberg}, we obtain to zero order in
$h$ and $\tilde{h}$:

\begin{eqnarray}
\bar{\delta}h_{\mu \nu} &=& c_{0 \mu ; \nu} + c_{0 \nu ; \mu} \\
\bar{\delta}\tilde{h}_{\mu \nu} &=& c_{1 \mu ; \nu} + c_{1 \nu ;
\mu} + \tilde{g}_{\mu \nu ; \lambda}c_0^{\lambda} + \tilde{g}_{\mu
\lambda}c_{0 ; \nu}^{\lambda} + \tilde{g}_{\nu \lambda}c_{0 ;
\mu}^{\lambda}
\end{eqnarray}

we also have:

\begin{eqnarray}
\bar{\delta} c_0^{\mu} &=& c_0^{\rho} c_{0, \rho}^{\mu} \nonumber \\
\bar{\delta} c_1^{\mu} &=& c_0^{\rho} c_{1, \rho}^{\mu} +
c_1^{\rho} c_{0,\rho}^{\mu} \nonumber \\
\bar{\delta} \bar{c}_0^{\mu} &=& i b^{\mu}_0 ( x ) \nonumber \\
\bar{\delta} \bar{c}_1^{\mu} &=& i b^{\mu}_1 ( x )
\end{eqnarray}

for the corresponding anti-ghosts $\bar{c}$ and where the $b$'s
are the auxiliary Nakanishi-Lautrup fields which satisfy:

\begin{eqnarray}
\bar{\delta} b_{0, 1}^{\mu} = 0
\end{eqnarray}

It has been verified that these transformations are nilpotent.
Now, we choose for our gauge fixing term:

\begin{eqnarray}
\label{GF grav} \textrm{GF} = -\sqrt{-g} \frac{C^2}{2} -
\tilde{\delta} \left( \kappa_2 \sqrt{-g} \frac{C^2}{2} \right)
\end{eqnarray}

we see that this is a good choice for our gauge fixing since it is
invariant under both transformations $\delta_0$ and $\delta_1$
(see \textbf{\ref{Sec: Mod. Theory Inv}}), where \cite{tHooft}
\cite{One Loop Div}:

\begin{eqnarray}
C^2 &=& g^{\alpha \beta} C_{\alpha} C_{\beta} \nonumber \\
C_{\mu} &=& D_{\nu} h_{\mu}^{\nu} - \frac{1}{2} D_{\mu}
h_{\nu}^{\nu}
\end{eqnarray}

In this way we have:

\begin{eqnarray}
\textrm{GF} &=& - \sqrt{-g} \left[ \left( 1 + \frac{\kappa_2}{2}
g^{\alpha \beta} \tilde{g}_{\alpha \beta} \right) \frac{C^2}{2} +
\kappa_2 \tilde{\delta} \left(\frac{g^{\mu \rho} C_{\mu}
C_{\rho}}{2}
\right) \right] \nonumber \\
&=& - \sqrt{-g} \left[ \left( 1 + \frac{\kappa_2}{2} g^{\alpha
\beta} \tilde{g}_{\alpha \beta} \right) \frac{C_{\mu} C^{\mu}}{2}
+ \kappa_2 \left(\tilde{C}_{\mu} C^{\mu} - \frac{\tilde{g}_{\mu
\beta} C^{\mu}C^{\beta}}{2} \right) \right]
\end{eqnarray}

where:

\begin{eqnarray}
\tilde{C}_{\mu} = \tilde{\delta} C_{\mu} = \tilde{\delta} \left[
D_{\nu}h_{\mu}^{\nu} - \frac{1}{2} D_{\mu} h_{\nu}^{\nu} \right] =
g^{\nu \rho}\left[ \nabla_{\nu} \tilde{h}_{\rho \mu} - \frac{1}{2}
\nabla_{\mu}\tilde{h}_{\rho \nu} \right] - \tilde{g}^{\nu \rho}
\left[ D_{\nu}h_{\rho \mu} - \frac{1}{2} D_{\mu} h_{\rho \nu}
\right]
\end{eqnarray}

this can be written in the form:

\begin{eqnarray}
\label{Gauge Fix.}
\textrm{GF} = - \sqrt{-g} H_{\mu} C^{\mu}
\end{eqnarray}

with:

\begin{eqnarray}
H_{\mu} = \left[ \left( 1 + \frac{\kappa_2}{2}
\tilde{g}^{\alpha}_{\alpha} \right) \frac{C_{\mu}}{2} + \kappa_2
\left( \tilde{C}_{\mu} - \tilde{g}_{\mu \beta} \frac{
C^{\beta}}{2} \right) \right]
\end{eqnarray}

Having established the form of the gauge fixing term we can now by
a standard procedure (the BRST method) find the associated Faddeev
Popov lagrangian. Following \cite{Kugo}, now we do:

\begin{eqnarray}
\mathcal{L}_{\textrm{GF} + \textrm{FP}} = - i \bar{\delta} ( P )
\end{eqnarray}

where $P$ in our case is:

\begin{eqnarray}
P = \bar{c}_0^{\mu} H_{\mu} + \bar{c}_1^{\mu} C_{\mu} + \beta_1
\bar{c}_1^{\mu} b_{0 \mu} + \beta_2 \bar{c}_0^{\mu} b_{1 \mu}
\end{eqnarray}

where the $\beta$'s are arbitrary constants to be fixed shortly,
so we have:

\begin{eqnarray}
\mathcal{L}_{\textrm{GF} + \textrm{FP}} = - i ( i b^{\mu}_0
H_{\mu} + i b^{\mu}_1 C_{\mu} + i ( \beta_1 + \beta_2 ) b^{\mu}_1
b_{0 \mu} - \bar{c}_0^{\mu} ( \bar{\delta} H_{\mu} ) -
\bar{c}_1^{\mu} ( \bar{\delta}C_{\mu} ) )
\end{eqnarray}

and so:

\begin{eqnarray}
\mathcal{L}_{\textrm{GF}} &=& b^{\mu}_0 H_{\mu} + b^{\mu}_1
C_{\mu} + ( \beta_1 + \beta_2 ) b^{\mu}_1 b_{0 \mu} \\
\label{FP grav} \mathcal{L}_{\textrm{FP}} &=& i ( \bar{c}_0^{\mu}
( \bar{\delta} H_{\mu} ) + \bar{c}_1^{\mu} ( \bar{\delta} C_{\mu}
) )
\end{eqnarray}

Now, for the gauge fixing part, we can use the equations of motion
for the auxiliary fields to make them disappear,

\begin{eqnarray}
\frac{\partial \mathcal{L}_{\textrm{GF}}}{\partial b^{\mu}_1} &=&
C_{\mu} + ( \beta_1 + \beta_2 ) b_{0 \mu} = 0 \longrightarrow b_{0
\mu} = -
\frac{C_{\mu}}{( \beta_1 + \beta_2 )} \nonumber \\
\frac{\partial \mathcal{L}_{\textrm{GF}}}{\partial b^{\mu}_0} &=&
H_{\mu} + ( \beta_1 + \beta_2 ) b_{1 \mu} = 0 \longrightarrow b_{1
\mu} = - \frac{H_{\mu}}{( \beta_1 + \beta_2 )}
\end{eqnarray}

substituting in $\mathcal{L}_{\textrm{GF}}$ we get:

\begin{eqnarray}
\mathcal{L}_{\textrm{GF}} = - \frac{C^{\mu} H_{\mu}}{( \beta_1 +
\beta_2 )} - \frac{C^{\mu} H_{\mu}}{( \beta_1 + \beta_2 )} +
\frac{( \beta_1 + \beta_2 ) C^{\mu} H_{\mu}}{( \beta_1 + \beta_2
)^2} = - \frac{C^{\mu} H_{\mu}}{( \beta_1 + \beta_2 )}
\end{eqnarray}

so we see we recover our initial gauge fixing if we set $( \beta_1
+ \beta_2 ) = 1$. Now for the Faddeev Popov lagrangian we have:

\begin{eqnarray}
\label{Faddeev Popov} \mathcal{L}_{\textrm{FP}} = i
\left(\bar{c}_0^{\mu} ( \bar{\delta} H_{\mu} ) + \bar{c}_1^{\mu} (
\bar{\delta} C_{\mu} )\right)
\end{eqnarray}

it is well known that \cite{tHooft} \cite{One Loop Div}:

\begin{eqnarray}
\bar{\delta} C_{\mu} = D_{\nu} D^{\nu} c_{0 \mu} + R_{\mu \nu}
c^{\nu}_0
\end{eqnarray}

and using:

\begin{eqnarray}
\bar{\delta} h_{\nu \rho} &=& D_{\nu} c_{0 \rho} + D_{\rho} c_{0
\nu} \nonumber \\
\bar{\delta} \tilde{h}_{\nu \rho} &=& \nabla_{\nu} c_{1 \rho} +
\nabla_{\rho} c_{1 \nu}
\end{eqnarray}

we get:

\begin{eqnarray}
\bar{\delta} H_{\mu} &=& \left[ \left( 1 + \frac{\kappa_2}{2}
\tilde{g}^{\alpha}_{\alpha} \right) \frac{\bar{\delta}C_{\mu}}{2}
+ \kappa_2 \left( \bar{\delta} \tilde{C}_{\mu} - \tilde{g}_{\mu
\beta} \frac{\bar{\delta} C^{\beta}}{2}
\right) \right] \nonumber \\
\bar{\delta} \tilde{C}_{\mu} &=& \nabla_{\nu} \nabla^{\nu} c_{1
\mu} + R_{\mu \nu} c^{\nu}_1 - g^{\rho \nu} \tilde{\delta}(
R^{\alpha}_{~ \rho \nu \mu} ) c_{0 \alpha} - \tilde{g}^{\nu \rho}
[ D_{\nu} D_{\rho} c_{0 \mu} + c_{0 \sigma} R^{\sigma}_{\rho \mu
\nu} ]
\end{eqnarray}

So, evaluating in (\ref{Faddeev Popov}), we will obtain (\ref{L
FP}).\\

\newpage

\section*{APPENDIX B: Background Field Method.}

The Background Field Method (BFM) is a mechanism utilized to
calculate the effective action at any order of perturbation theory
without losing explicit gauge invariance. This simplifies the
calculations and the comprehension of the model. The importance of
the Effective Action is due to the fact that it contains all the
quantum information of the theory and that from it all
One-Particle-Irreducible (1PI) Feynman diagrams can be computed.
Stringing them together, we can compute all connected Feynman
diagrams in a more efficient manner \cite{Abbott} and from them
the S- matrix can be calculated.\\

Next we calculate the effective action $\Gamma$ for a general
model using the BFM. One begins by defining the generating
functional of disconnected diagrams $Z[J]$:

\begin{eqnarray}
Z[J] = \int \mathcal{D} \varphi e^{i(S[\varphi] + J \cdot
\varphi)}
\end{eqnarray}

where $S$ is the action of the system and where we will be using
the notation $J \cdot \varphi \equiv \int J \varphi d^4 x$. In the
background field method we do the identification $\varphi
\rightarrow \varphi + \phi$ inside the action, where $\phi$ is an
arbitrary background. So now we have:

\begin{eqnarray}
\hat{Z}[J,\phi] = \int \mathcal{D} \varphi e^{i(S[\varphi + \phi]
+ J \cdot \varphi)}
\end{eqnarray}

Now the generating functional of connected diagrams $W[J]$ is:

\begin{eqnarray}
W[J] = - i \ln Z[J]
\end{eqnarray}

so we define:

\begin{eqnarray}
\hat{W}[J,\phi] = - i \ln \hat{Z}[J,\phi]
\end{eqnarray}

and

\begin{eqnarray}
\bar{\varphi} = \frac{\delta W}{\delta J}
\end{eqnarray}

so here

\begin{eqnarray}
\hat{\varphi} = \frac{\delta \hat{W}}{\delta J}
\end{eqnarray}

with all these definitions it is possible to give the formula for
the usual Effective Action:

\begin{eqnarray}
\Gamma[\bar{\varphi}] = W[J] - J \cdot \bar{\varphi}
\end{eqnarray}

and the background field Effective Action:

\begin{eqnarray}
\hat{\Gamma}[\hat{\varphi},\phi] = \hat{W}[J,\phi] - J \cdot
\hat{\varphi}
\end{eqnarray}

now we do the shift $\varphi \rightarrow \varphi - \phi$ so that:

\begin{eqnarray}
\hat{Z}[J,\phi] = Z[J] e^{- i J \cdot \phi}
\end{eqnarray}

from which it follows (after taking logarithms):

\begin{eqnarray}
\hat{W}[J,\phi] = W[J] - J \cdot \phi
\end{eqnarray}

taking now the functional derivative with respect to $J$:

\begin{eqnarray}
\hat{\varphi} = \bar{\varphi} - \phi
\end{eqnarray}

but now we can appreciate that:

\begin{eqnarray}
\hat{\Gamma}[\hat{\varphi},\phi] &=& W[J] - J \cdot \phi - J
\cdot \hat{\varphi} \nonumber \\
&=& W[J] - J \cdot \phi - J \cdot (\bar{\varphi} - \phi )
\nonumber
\\
&=& W[J] - J \cdot \bar{\varphi} \nonumber \\
\hat{\Gamma}[\hat{\varphi},\phi] &=& \Gamma[\hat{\varphi} + \phi]
\end{eqnarray}

In particular if we take $\hat{\varphi} = 0$, we have:

\begin{eqnarray}
\hat{\Gamma} [0,\phi] = \Gamma[\phi]
\end{eqnarray}

This means that the Effective Action of the theory $\Gamma$ can be
computed from the background field Effective Action $\hat{\Gamma}$
by taking the quantum field to zero and with the presence of the
background $\phi$. Since the derivatives of the Effective Action
with respect to the fields generate the 1PI diagrams, the last
equation means that if we treat $\phi$ perturbatively what we will
have will be diagrams with external legs corresponding to the
background field $\phi$ and with internal lines corresponding to
the quantum field $\varphi$.\\

And so, to study the quantum effects it only suffices to do an
expansion in the quantum fields in the action $S$ or in the
lagrangian $L$ using the identification of the Background Field
Method. This means:

\begin{eqnarray}
\label{BFM} \phi_I \rightarrow \phi_I + \varphi_I
\nonumber \\
\tilde{\phi}_I \rightarrow \tilde{\phi}_I + \tilde{\varphi}_I
\end{eqnarray}

We use (\ref{BFM}) in $\tilde{\delta}$ Gravity, where $g_{\mu \nu}
\rightarrow g_{\mu \nu} + h_{\mu \nu}$ and $\tilde{g}_{\mu \nu}
\rightarrow \tilde{g}_{\mu \nu} + \tilde{h}_{\mu \nu}$.\\

\newpage

\section*{APPENDIX C: Divergent Part of the Effective Action at One Loop.}

As was mentioned in \textbf{Section \ref{Chap: Quant correc Mod
Theo}} there are various ways to calculate the divergent part of
the effective action at one loop, but they are quite complicated.
So, we have resolved to follow an algorithm developed in \cite{One
Loop Div}.\\

The Effective Action $\Gamma$ to one loop can be written as:

\begin{eqnarray}
\Gamma[\phi] = S[\phi] + \frac{i}{2} \hbar \textrm{Tr} \ln D +
O(\hbar^2)
\end{eqnarray}

where:

\begin{eqnarray}
D_i^{~j} = \frac{\delta^2 S}{\delta \phi_i \delta \phi_j}[\phi]
\end{eqnarray}

is a differential operator depending on the background field
$\phi_i$. Its most general form is:

\begin{eqnarray}\label{D} D_i^{~j} &=& K^{\mu_1 \mu_2 \ldots \mu_L ~ j}_{ ~~~~~~~~~~ i} \nabla_{\mu_1} \nabla_{\mu_2} \ldots \nabla_{\mu_L} + S^{\mu_1 \mu_2 \ldots \mu_{L - 1} ~ j}_{ ~~~~~~~~~~~~~ i} \nabla_{\mu_1} \nabla_{\mu_2} \ldots \nabla_{\mu_{L - 1}} \nonumber \\
&+& W^{\mu_1 \mu_2 \ldots \mu_{L - 2} ~ j}_{ ~~~~~~~~~~~~~ i}
\nabla_{\mu_1} \nabla_{\mu_2} \ldots \nabla_{\mu_{L - 2}} +
N^{\mu_1 \mu_2 \ldots \mu_{L - 3} ~ j}_{ ~~~~~~~~~~~~~ i}
\nabla_{\mu_1} \nabla_{\mu_2} \ldots
\nabla_{\mu_{L - 3}} \nonumber \\
&+& M^{\mu_1 \mu_2 \ldots \mu_{L - 4} ~ j}_{ ~~~~~~~~~~~~~ i}
\nabla_{\mu_1} \nabla_{\mu_2} \ldots \nabla_{\mu_{L - 4}} + \ldots
\end{eqnarray}

where $K, S, W, N, M$ are parameters which must be specify for
each model and $\nabla_{\mu}$ is a covariant derivative:

\begin{eqnarray}
\nabla_{\alpha} T^{\beta ~ j}_{~ i} &=& \partial_{\alpha} T^{\beta
~ j}_{~ i} + \Gamma_{\alpha \gamma}^{~~ \beta} T^{\gamma ~ j}_{~
i} + \omega_{\alpha ~ i}^{~ k} T^{\beta ~ j}_{~ k} - \omega_{\alpha ~ k}^{~ j} T^{\beta ~ k}_{~ i} \\
\nabla_{\mu} \Phi_i &=& \partial_{\mu} \Phi_i + \omega_{\mu ~
i}^{~ j} \Phi_j
\end{eqnarray}

here:

\begin{eqnarray}
\Gamma^{~~ \alpha}_{\mu \nu} = \frac{1}{2} g^{\alpha \beta}
(\partial_{\mu} g_{\nu \beta} + \partial_{\nu} g_{\mu \beta} -
\partial_{\beta} g_{\mu \nu})
\end{eqnarray}

and $\omega_{\mu ~ i}^{~ j}$ is a connection on the principle
bundle. The computation of the divergent part of Effective Action
at one loop is done through a lengthy and cumbersome calculation
that consist in the sum of a finite number of one loop divergent
Feynman diagrams, the details are given in \cite{One Loop Div} and
the result by equation (30) in the same reference. This last
result is too large to show here, but it depends on the parameters
involved in $D_i^{~j}$ (\ref{D}). So basically, what we need is
the quadratic part of the lagrangian of the model to obtain the
divergent part of the effective action.\\

In $\tilde{\delta}$ Gravity, we have:

\begin{eqnarray}
\phi_i \rightarrow \vec{h}_{( \alpha \beta )}
\end{eqnarray}

where $\vec{h}$ is defined in (\ref{h vect}). As the covariant
derivative acting on $\vec{h}$ is given by (\ref{Cov h}) this
means $i \rightarrow ( \alpha \beta )$:

\begin{eqnarray}
\omega_{\mu ~ i}^{~ j} \rightarrow - \left([\Gamma_{\mu
\alpha}^{~~ \rho}] \delta_{\beta}^{\nu} + [\Gamma_{\mu \beta}^{~~
\rho}] \delta_{\alpha}^{\nu}\right)
\end{eqnarray}

where $[ \Gamma_{\mu \alpha}^{~~ \rho} ]$ is given by equation
(\ref{Matrix Gamma}). The other relevant parameters in our model
are given by:\\

$L = 2$\\

$K^{\mu_1 \mu_2 \ldots \mu_L ~ j}_{ ~~~~~~~~~~ i}$ given by (\ref{K grav}).\\

$S^{\mu_1 \mu_2 \ldots \mu_{L - 1} ~ j}_{ ~~~~~~~~~~~~~ i} = 0$.\\

$W^{\mu_1 \mu_2 \ldots \mu_{L - 2} ~ j}_{ ~~~~~~~~~~~~~ i}$ given by (\ref{W grav}).\\

On the other side, to the Faddeev Popov ghosts we have:

\begin{eqnarray}
\phi_i \rightarrow \vec{c}_{\alpha}
\end{eqnarray}

where $\vec{c}_{\alpha}$ is defined by (\ref{c vect}) and the
Covariant Derivative (\ref{Cov ghost}) says us that:

\begin{eqnarray}
\omega_{\mu ~ i}^{~ j} \rightarrow - [\Gamma_{\mu \alpha}^{~~
\rho}]
\end{eqnarray}

Finally, the other parameters are given by:\\

$L = 2$\\

$K^{\mu_1 \mu_2 \ldots \mu_L ~ j}_{ ~~~~~~~~~~ i}$ given by (\ref{K fant}).\\

$S^{\mu_1 \mu_2 \ldots \mu_{L - 1} ~ j}_{ ~~~~~~~~~~~~~ i} = 0$.\\

$W^{\mu_1 \mu_2 \ldots \mu_{L - 2} ~ j}_{ ~~~~~~~~~~~~~ i}$ given by (\ref{W fant}).\\


\newpage


\begin{thebibliography}{99}

\bibitem{GR}S. Weinberg. \textit{Gravitation and Cosmology: Principles and Applications of the General Theory of Relativity.}
John Wiley and Sons, Inc. 1 edition, 1972; C.W. Misner, K.S.
Thorne, J.A. Wheeler. \textit{Gravitation.} W.H. Freeman and
Company, twenty third printing 2000.

\bibitem{SolSist} C. M. Will, \textit{The Confrontation between General Relativity and Experiment}, Living Rev. Relativity 9, (2006),
http://www.livingreviews.org/lrr-2006-3.

\bibitem{tHooft} G. 'tHooft and M. Veltman. Annales de l'I.H.P. Section A, tome $20$ (1974), page $69$-$94$.

\bibitem{DeWitt} B. DeWitt. Physical Review. Vol $160$ $(1967)$, page $1113$-$1148$.

\bibitem{String 1} M.B. Green, J.H. Schwarz and E. Witten.
\textit{Super String Theory.} Vols 1,2. Cambridge University Press
$1987$.

\bibitem{String 2} J. Polchinski. \textit{Super String
Theory.} Vols 1,2. Cambridge University Press $1998$.

\bibitem{LQG} See for instance, C. Rovelli. \textit{Quantum Gravity.} Cambridge
University Press $2007$, and references therein.

\bibitem{Twistors} R. Penrose and W. Rindler. \textit{Spinor and Space-Time, Spinor and Twistor Method in Space-Time
Geometry.} Vol 2. Cambridge University Press $1988$.

\bibitem{non-commutative} A. Connes. \textit{Noncommutative
Geometry.} Academic Press $1994$.

\bibitem{LandScape} M. R. Douglas. JHEP05(2003)046; J. Kumar. International Journal of Modern Physics
A. Vol 21 (2006), page 3441–3472, and references therein.

\bibitem{bhe2}See J. F. Barbero and E. Villase\~nor. Classical Quantum
Gravity. Vol 26 (2009), 035017. And references therein.

\bibitem{Ashtekar} A. Ashtekar, L. Bombelli and A. Corichi. Physical Review D. Vol 72 (2005),
025008.

\bibitem{weias}S. Weinberg. \textit{General Relativity: An Einstein
centenary survey}. Edited by S. W.Hawking and W.Israel. Cambridge
University Press, 1979. Chapter 16, page 790.

\bibitem{adler}Ya.B. Zeldovich, JETP Lett., 6, 316 (1967); A. Sakharov, SOv. Phys. Dokl., 12, 1040 (1968);
O. Klein, Phys. Scr. 9, 69 (1974); S. Adler, Rev. Mod. Phys., 54,
729 (1982).

\bibitem{litim}D.F. Litim, Phys.Rev.Lett.92:201301,2004; AIP Conf. Proc. 841, 322 (2006);
e-Print: arXiv:0810.367; A. Codello, R. Percacci and C. Rahmede,
Annals Phys.324:414-469,2009; M. Reuter and F. Saueressig,
Lectures given at First Quantum Geometry and Quantum Gravity
School, Zakopane, Poland (2007),arXiv:0708.1317

\bibitem{loll}J. Ambjorn, J. Jurkiewicz and R. Loll. Physical Review Letters. Vol 85 (2000), page $924-927$.

\bibitem{aep} J. Alfaro, D. Espriu and D. Puigdomenech. Physical Review D. Vol 82 (2010), 045018.

\bibitem{salam}C.J. Isham, A. Salam and J.A. Strathdee. Annals Phys. Vol 62 (1971), page $98-119$.

\bibitem{ogi}A.B. Borisov and V.I. Ogievetsky. Theor.Math.Phys.21:1179,1975;
E.A. Ivanov and V.I. Ogievetsky. Lett.Math.Phys.1:309-313,1976.

\bibitem{ru}D. Amati and J. Russo. Physics Letters B. Vol 248 (1990), page $44-50$;
J. Russo. Physics Letters B. Vol 254 (1991), page $61-65$; A.
Hebecker, C. Wetterich. Physics Letters B. Vol 574 (2003), page
$269-275$; C. Wetterich. Physical Review D. Vol 70 (2004), 105004.

\bibitem{One Loop Div} Petr I. Pronin and Konstantin V. Stepanyantz. Nuclear Physics B. Vol $485$ ($1997$),
page $517$-$544$.

\bibitem{Non-Renormalizable 1} Marc H. Goroff and Augusto Sagnotti. Physics Letters B. Vol $160$ (1985), page
$81$-$86$.

\bibitem{Non-Renormalizable 2} Marc H. Goroff and Augusto
Sagnotti. Nuclear Physics B. Vol $266$ ($1986$), page $709$-$736$.

\bibitem{Non-Renormalizable 3} Anton E.M. van de Ven. Nuclear Physics B. Vol $378$ ($1992$), page $309$-$366$.

\bibitem{Yang Mills} J. Alfaro, bv gauge theories, hep-th 9702060;
Jorge Alfaro and Pedro Labra\~na. Physical Review D. Vol $65$
(2002), $045002$.

\bibitem{JA} J. Alfaro. arXiv:1006.5765v1 [gr-qc]. June 30, 2010.

\bibitem{phantom 1} R.R. Caldwell. Physics Letters B. Vol 545 (2002), page 23-29.

\bibitem{phantom 2} R. R. Caldwell, Marc Kamionkowski and Nevin N.Weinberg. Physical Review Letters. Vol 91 (2003), 071301.

\bibitem{phantom 3} S. Nojiri and S. D. Odintsov. Physics Letters B. Vol 562 (2003),
page 147-152.

\bibitem{phantom 4} J. M. Cline, S. Jeon and G. D. Moore. Physical Review D. Vol 70 (2004),
043543.

\bibitem{phantom 5} G. W. Gibbons. arXiv:hep-th/0302199v1 (2008).

\bibitem{SG}  M. T. Grisaru, P. van Nieuwenhuizen and J. A. M. Vermaseren. Physical Review Letters. Vol 37 (1976), page $1662$-$1666$.

\bibitem{cubic term} Ming Lu and Mark B. Wise. Physical Review D. Third series, Vol $47$ (1993), page R3095-R3098.

\bibitem{Dobado1} A. Dobado and A. Lopez. Physics Letters B. Vol $316$ ($1993$), page $250$-$256$.

\bibitem{Dobado2} A. Dobado and A. L. Maroto. Physical Review D. Vol $52$ (1995), page 1895-1901.

\bibitem{Non local} J.A. Cabrer and D. Espriu. Physics Letters B. Vol $663$ ($2008$), page 361-366.

\bibitem{Kugo} T. Kugo and S. Uehara. Nuclear Physics B. Vol $197$ ($1982$), page $378$-$384$.

\bibitem{Abbott} L. F. Abbott. Acta Physica Polonica B. Vol 13 (1982), page 33-50.

\bibitem{Abers Lee} E.S. Abers and B.W. Lee. \textit{Gauge Theories.} Physics Reports $9$, No $1$, page 99.

\bibitem{ramond} P. Ramond. \textit{Field Theory: A Modern Primer.} The Benjamin/Cummings Publishing Company,
INC. $1981$. Section 3.4.

\bibitem{ZReg} P. Ramond. \textit{Field Theory: A Modern Primer.} The Benjamin/Cummings Publishing Company,
INC. $1981$. page 115.

\bibitem{FORM} J.A.M.Vermaseren \textit{New features of FORM.} math-ph/0010025.

\bibitem{NCGI} I.I.Shapiro, W.B. Smith, M.B. Ash, R.P. Ingalls and
G.H. Pettengill, Phys. Rev. Lett. Vol 26 (1971), page 27-30.

\bibitem{NCGII} E. Gaztanaga, E. Garcia-Berro, J. Isern,
E. Bravo and I. Dominguez, Phys. Rev. D. Vol 65 (2001), 023506.

\bibitem{GMTW} C.W. Misner, K.S. Thorne, J.A. Wheeler.
\textit{Gravitation.} W.H. Freeman and Company, twenty third
printing 2000, page 180.

\bibitem{Kugo 2} T. Kugo and I. Ojima. \textit{Supplement of the Progress
of Theoretical Physics}, No 66 (1979). See Chapter III.

\bibitem{SusyHam1} E. Witten, \textit{Lectures Notes on Supersymmetry}, Trieste Lectures. July,
1981.

\bibitem{SusyHam2} S. J. Gates Jr., M.T. Grisaru, M. Rocek, W. Siegel, \textit{Superspace
or One Thousand and One Lessons in Supersymmetry}, The Benjamin
Cumming Publishing Company. page 65.

\bibitem{Weinberg} S. Weinberg. \textit{The Quantum Theory Fields.}
Vol II, Cambridge University Press 1996, page 97.

\end{thebibliography}
\end{document}